\documentclass[iop,jphysa,reprint,superscriptaddress,longbibliography,eprint]{revtex4-1} 

\usepackage[utf8]{inputenc}

\makeatletter\def\paragraph{%
  \@startsection{paragraph}{4}{\parindent}%
  {\z@}%
  {-1em}%
  {\reset@font\normalsize\itshape}%
}\makeatother%

\usepackage{cmap} 
\usepackage[T1]{fontenc}
\usepackage[normalem]{ulem}

\usepackage{amsmath,amssymb,amsfonts,amsthm,mathtools,mleftright}
\usepackage{braket}
\usepackage{lmodern}
\usepackage{graphicx}
\usepackage[dvipsnames]{xcolor}
\usepackage{ulem}
\usepackage{hyperref}
\hypersetup{colorlinks=true, linkcolor=BrickRed,
  urlcolor=blue!50!black, citecolor=blue!50!black}

\usepackage{cleveref}
\crefrangelabelformat{equation}{(#3#1#4)$-$(#5#2#6)}
\crefname{eqset}{eqs.}{eqs.} 
\Crefname{eqset}{Eqs.}{Eqs.}
\crefformat{eqset}{eqs.~(#2#1#3)}
\Crefformat{eqset}{Eqs.~(#2#1#3)}
\crefmultiformat{eqset}{eqs.~(#2#1#3)}{ and~(#2#1#3)}{, (#2#1#3)}{ and~(#2#1#3)}

\newcommand{\rev}[1]{#1}
\newcommand{\rrev}[1]{#1}

\newcommand{\FunctionSpace}{H}
\newcommand{\Fockspace}{F(\FunctionSpace)}
\newcommand{\SpaceOfMotion}{\mathbb{X}}
\newcommand{\SymmOperator}{\mathcal{S}}
\newcommand{\ReactionOperator}{\mathcal{R}}
\newcommand{\AltReactionOperator}{\mathcal{Q}}
\newcommand{\Propensity}{\lambda}

\newcommand{\aps}[1]{a^+\{ #1 \}}
\newcommand{\ams}[1]{a^-\{ #1 \}}

\newcommand{\rhovac}{\rho_\text{vac}}
\newcommand{\Species}{\text{A}}

\DeclareMathOperator\Probability{\mathbb{P}}
\DeclareMathOperator\Expectation{\mathbb{E}}
\newcommand\Prob[1]{\Probability\mleft[#1\mright]}
\newcommand\Expect[1]{\Expectation\mleft[#1\mright]}
\DeclareMathOperator\Id{\mathcal{I}}
\DeclareMathOperator\Span{span}
\DeclareMathOperator\dom{dom}

\renewcommand\i{\mathrm{i}}
\renewcommand\leq{\leqslant}
\renewcommand\geq{\geqslant}
\renewcommand\emptyset{\varnothing}
\renewcommand\epsilon{\varepsilon}

\begin{document}

\title{A probabilistic framework for particle-based reaction-diffusion dynamics using classical Fock space representations 
\footnote{To appear in Letters in Mathematical Physics on the themed collection: Mathematical Physics and Numerical Simulation of Many-Particle Systems; V. Bach and L. Delle Site (eds.)}
}

\newcommand\ZIB{\affiliation{Zuse Institute Berlin, Takustr. 9, 14195 Berlin, Germany}}
\newcommand\FUB{\affiliation{Freie Universität Berlin, Department of Mathematics and Computer Science, Arnimallee 6, 14195 Berlin, Germany}}
\newcommand\UVA{\affiliation{Van ’t Hoff Institute for Molecular Sciences and Korteweg-de Vries Institute for Mathematics, University of Amsterdam, 1090GD and 1090GE Amsterdam,
The Netherlands}\affiliation{Dutch Institute for Emergent Phenomena, 1090GL Amsterdam, The Netherlands
}}

\author{Mauricio J. del Razo}\email{m.delrazo@fu-berlin.de}
\UVA
\FUB

\author{Daniela Frömberg}
\FUB

\author{Arthur V. Straube}
\ZIB

\author{Christof Schütte}
\ZIB \FUB

\author{Felix Höfling}
\FUB \ZIB

\author{Stefanie Winkelmann}
\ZIB

\date{\today}
\begin{abstract}	
The modeling and simulation of stochastic reaction-diffusion processes is a topic of steady interest that is approached with a wide range of methods. \rev{At the level of particle-resolved descriptions,
where chemical reactions are coupled to the spatial diffusion of individual particles, there exist comprehensive numerical simulation schemes, while the corresponding mathematical formalization is relatively underdeveloped.
The aim of this paper is to provide a framework to systematically formulate the probabilistic evolution equation, termed \textit{chemical diffusion master equation} (CDME), that governs particle-based stochastic reaction-diffusion processes.}
To account for the non-conserved and unbounded particle number of this type of open systems, we employ a classical analogue of the quantum mechanical Fock space that contains the symmetrized probability densities of the many-particle configurations in space.
Following field-theoretical ideas of second quantization, we introduce creation and annihilation operators that act on single-particle \rev{densities} and provide natural representations of symmetrized probability densities as well as of reaction and diffusion operators. \rev{These operators allow us to consistently and systematically formulate the CDME for arbitrary reaction schemes. The resulting form of the CDME further} serves as the foundation to derive more coarse-grained descriptions of reaction-diffusion dynamics. In this regard, we show that a discretization of the evolution equation by projection onto a Fock subspace generated by a finite set of single-particle densities leads to a generalized form of the well-known reaction-diffusion master equation, which supports non-local reactions between grid cells and which converges properly in the continuum limit.


\end{abstract}

\maketitle

\section{Introduction}

A great variety of chemical and biochemical phenomena on all scales hinge on the combination of diffusion and chemical reactions;
examples range from classical front propagation \cite{Fisher1937frontprop, KPP1937frontprop, vanSaarloos2003frontprop},
self-organization of excitable media\cite{Nicolis1977book, Kuramoto1984book,Vasquez2004reactchannel, Straube2007reactmicro} 
and pattern-forming coatings of animals \cite{Cross1993pattform, murray2003book}, 
over the formation of morphogen gradients \cite{wartlick2009morphgrad, fradin2017morphgrad, huang2020morphgrad, stapornwongkul2021morphgrad}
and MinE protein oscillations\cite{halatek2012mine, amiranashvili2016mine, denk2018mine} in developmental biology, 
to 
spreading of diseases \cite{britton2019epidemic, sego2021generation, ganyani2021simulation} and innovations \cite{djurdjevac2018human}.
Conventionally, such processes have been modelled by means of deterministic reaction-diffusion equations, which govern the temporal evolution of concentration fields \cite{epstein-pojman1998book, murray2003book}.
Such continuum descriptions, however, often break down for small copy numbers of molecules, as is typically the case for biochemical processes at cellular and subcellular scales \cite{grima2008intracell, wilkinson2009stochmod, smith2018spatial}.
The discreteness of copy numbers gives rise to intrinsic noise, which has been included heuristically in reaction-diffusion models in the form of spatiotemporal Gaussian white noise, e.g., in calcium signalling \cite{falcke2003calcium, ruediger2014calcium, powell2019calcium, friedhoff2021calcium}, but also for species extinction \cite{reichenbach2007noise, ovaskainen2010extinction}.

A more systematic, microscopic approach suggests to switch from concentration fields to spatially-resolved probability distributions of copy numbers, which leads to the reaction-diffusion master equation (RDME) \cite{hellander2015reaction,isaacson2009reaction,isaacson2006incorporating}: \rev{position} space is partitioned into a regular mesh of volume elements, the diffusion of molecules is replaced by a jump process on the mesh, and reactions occur only within each volume element under assumed well-mixed conditions.
Such conditions are justified in the spatio-temporal chemical master equation (ST-CME) \cite{winkelmann2016spatiotemporal,winkelmann2020stochastic}, which has been put forward in the context of cellular environments to account for intracellular structures and compartmentalization. It uses an irregular partitioning of \rev{position} space into comparably few metastable subsets, meaning that the subsets are separated by barriers such that jumps between them occur rarely.
Another obstacle towards effective continuum models for biochemical processes is macromolecular crowding, i.e., the dense and heterogeneous packing of cellular spaces by macromolecules that do not participate in the reaction, which has consequences for product formation rates as well as diffusion-influenced reaction kinetics \cite{zhou2008macromolecular, hoefling2013anomalous, weiss2014crowding, lanoiselee2018diffusion, froemberg2008stationary, sereshki2012subdiffusion}.
Some of the aspects of crowding have been included in recent extensions of numerical reaction--diffusion schemes \cite{engblom2018mesoscopic, sarkar2020concentration}, showing the potential for qualitative changes of the observed phenomenology.

An alternative to the above probabilistic descriptions is stochastic simulations of particle-based reaction-diffusion (PBRD) models, which offer a high resolution down to the scale of molecules combined with great modelling flexibility.
The idea is that molecules are represented by point particles undergoing Brownian motion and that bi-molecular reactions between close-by particles occur with a given rate depending on the separation distance of the pair; the most common schemes use either a reaction volume (Doi model \cite{doi1976second,doi1976stochastic}) or a reaction surface (Smoluchowski model \cite{smoluchowski1917attempt}) in terms of this distance.
PBRD schemes are constructed in a bottom-up way and based on the extensive theory of diffusion-influenced reactions.\cite{agmon1990theory,hanggi1990reaction,rice1985diffusion,szabo1980first,szabo1982stochastically}
A number of algorithms for the PBRD scheme exist \cite{andrews2004stochastic, van2005green, erban2009stochastic, lipkova2011analysis, klann2012spatial, hoffmann2019readdy}, 
differing in their implementation of Brownian motion, whether physical interactions between molecules are supported \cite{dibak2019diffusion,froehner2018reversible}, and in their degree of molecular resolution \cite{delrazo2021multiscale, dibak2018msm, delrazo2018grandcanonical}.
Applications reach from enzyme kinetics under crowding conditions \cite{ridgway2008coarsegrained, echeverria2015enzyme, weilandt2019particlebased} to nanomaterial-based catalysis \cite{lin2020coverage}.
A recent comparison of the two probabilistic models given by RDME and 
PBRD with a focus on intracellular kinetics can be found in the review by \citet{smith2018spatial}.

A mathematical formalization of PBRD models as an open many-particle system undergoing reactions and diffusion was developed only partially \cite{doi1976second, doi1976stochastic, grassberger1980fock, birch2006master, dodd2009many, kolokoltsov2010nonlinear}. 
Yet, such a framework would be a highly desirable starting point to systematically derive numerical schemes for reaction-diffusion processes and to analytically connect PBRD models with coarse-grained descriptions such as the  RDME and the ST-CME.
For example, these compartmentalized descriptions are obtained by locally integrating out the spatial degrees of freedom, turning the diffusion process into a continuous-time random walk on a mesh of subdomains.
For bi-molecular reactions, however, the procedure gives rise to effective reaction rates \cite{gopich2013diffusion, gopich2019diffusion, dibak2019diffusion} and markedly non-Markovian reaction time distributions \cite{gopich2018theory, grebenkov2018strong,froemberg2021generalized}.
For the standard RDME, where second-order reactions may occur only between particles of the same subdomain, it was shown that for decreasing mesh size, the dynamics converge to a limit where second-order reactions cease to occur \cite{hellander2012reaction}.  
This problem was addressed in terms of a \textit{convergent RDME} \cite{isaacson2013convergent,isaacson2018unstructured}, which allows bimolecular reactions to take place also between particles located in different subdomains.

\rev{The aim of this work is to develop a framework for the systematic formulation of the probabilistic evolution equation for particle-based stochastic reaction-diffusion processes}, which we refer to as \textit{chemical diffusion master equation} (CDME).
The CDME is a family of Fokker-Planck equations, each of them describing the diffusion processes for a given $n$-particle probability density. The equations within the family are coupled by the reaction dynamics similar in form to a chemical master equation (CME) \cite{gillespie1976general, vanKampen1992stochastic}.
\rev{The underlying reaction schemes are not restricted to mass-conserving reactions and may include effective reactions such as insertion and removal of particles, e.g., by considering only a subset of the chemical species. Thus, the CDME is capable of describing the dynamics of an open system characterized by the overall number of particles changing in time, similarly to the standard CME \cite{qian2006open}. To systematically formulate the CDME given a general set of chemical reactions,} our framework uses creation and annihilation operators.
We exemplify the framework for a birth-death process as well as the second-order reaction of mutual annihilation.
Our work further provides the basis for putting reaction-diffusion models at different resolutions on the same footing, which clears the way to derive relationships between the various coarse-grained descriptions and numerical schemes for reaction-diffusion dynamics. \rev{The latter is fundamental to develop consistent multiscale simulations.}
As a first step, we discretize the particle-based dynamics by doing a Galerkin projection \cite{deuflhard2008adaptive} of the CDME onto a partition of the \rev{position} space into subvolumes. This yields a generalized RDME, where bi-molecular reactions can occur naturally between particles in different, yet close-by subvolumes.
In particular, we find explicit relations between the reaction rate constants of the RDME and of the underlying particle-based model as given by the CDME. 

A number of technical challenges need to be solved to achieve these goals. In general, reaction-diffusion systems are open systems in the sense that the number of particles is not fixed, but changes in the course of time due to the reactions.
Whereas such situations are well-known in quantum field theory and solid state physics, where they are addressed in terms of creation and annihilation operators, the analogous formulations for classical systems of indistinguishable particles are comparably underdeveloped;
this is particularly true for the stochastic dynamics of particles undergoing Brownian motion in an open system.
In a probabilistic description of reaction-diffusion dynamics, the number of particles of each species is no longer determined but obeys a statistical distribution, which evolves in time along with the particle positions.
\rev{Further, the many-particle distributions do not distinguish individual particles of the same species, and the distributions are thus symmetric with respect to permutations of particle labels. The symmetry must be preserved under time evolution, which requires non-trivial combinatorial factors in the reaction operators. These aspects}
are addressed by borrowing the concept of the Fock space \cite{fock1932konfigurationsraum} from quantum mechanics and translating it to the classical setting
\cite{doi1976second,grassberger1980fock,bressloff2021construction}.
A major mathematical obstacle in this endeavor is that probability densities are integrable, but not necessarily square-integrable functions, as opposed to quantum wavefunctions, and that the underlying \rev{function} space is not a Hilbert space \cite{kolokoltsov2010nonlinear}. \rrev{Some of these issues have been addressed in the literature of probability theory for population dynamics \cite{birch2006master, carinci2015dualities, dodd2009many, jansen2014notion, kolokoltsov2010nonlinear}; however these works do not explore the applications in the context of chemical reactions.}

\paragraph*{Outline.}
We start in \cref{sec:CDME} by introducing a probabilistic model for particle-based reaction-diffusion dynamics and motivating the structure of the chemical diffusion-master equation that we look for. In \cref{sec:fockspace} the probability functions are interpreted as elements of a Fock space; the creation and annihilation operators are introduced and central algebraic relations are derived. \rev{The main results of this work are presented in \cref{sec:rep_operators}, where the creation and annihilation operators are used to express the symmetrized densities and the diffusion and reaction operators. This allows us to systematically formulate the CDME on a general level, as well as for two exemplary reaction systems.} Finally, Galerkin projection methods are applied in \cref{Spatial_dis} to derive the generalized RDME as a projection of the underlying CDME.

\rev{\section{Probabilistic model for particle-based reaction-diffusion dynamics}}\label{sec:CDME}

The object of interest in this work is a collection of molecules that diffuse in space and can undergo chemical reactions. These reactions will cause changes in the number of molecules of a given species.
Focusing on a single species as a first step, elementary reactions include the creation and annihilation of molecules,
which is then straightforward to extend to, e.g., binary reactions amongst molecules of different species.
The goal is to find a probabilistic description that includes both the spatial movement of molecules and the changes in the population size. 
\rev{In this section, we introduce the structure of the CDME and formulate the equations for an exemplary birth-death process. A systematic approach to formulate the CDME for general reaction schemes will be given in \cref{sec:rep_operators}.}

\subsection{\rev{Symmetric} probability density functions} \label{sec:probdens}

Specifically, we consider point particles of only one chemical species in a bounded domain $\SpaceOfMotion \subset \mathbb{R}^3$.
Due to the reactions, the number of particles $N(t) \in \mathbb{N}_0$ at time $t$ can vary in the course of time.
Assuming that there are $n=N(t)$ particles at time $t$, the configuration of all particle positions is denoted by the vector $x^{(n)}=\mleft(x^{(n)}_1,\dots,x^{(n)}_n\mright)\in \SpaceOfMotion^n$.
The statistical law of the configurations is encoded in the conditional probability density function $p_n(x^{(n)},t|N(t)=n)$ at time $t$ given that $N(t)=n$. Following the situation of a fixed particle number, we adopt the convention that $p_n$ is normalized:
\begin{equation}
	\int_{\SpaceOfMotion^n} p_n(x^{(n)},t|N(t)=n)  \, dx^{(n)} =1
\end{equation}
for all $t \geq 0$ and each fixed $n$.
The unconditional probability density for $n\geq 1 $ is then given by
\begin{equation} \label{denistydef}
	\rho_n(x^{(n)},t) := p_n(x^{(n)},t|N(t)=n) \Prob{N(t)=n} \,,
\end{equation}
where $\Prob{N(t)=n}$ is the probability of having $n$ particles in the system at time $t$;
for the empty system, we set $\rho_0(t) := \Prob{N(t) =0}$.
It follows that
\begin{equation} \label{P(N)}
	\Prob{N(t)=n} = \int_{\SpaceOfMotion^n} \rho_n (x^{(n)},t) \,dx^{(n)} \,.
\end{equation}
In particular, the admissible probability densities $\rho_n$ belong to the space of absolutely integrable functions,
$L^1(\SpaceOfMotion^n) = \{ \eta : \SpaceOfMotion^n \to \mathbb{R} \,; \|\eta\|_n < \infty \}$ with the standard norm
\begin{equation} \label{n-norm}
	\|\eta\|_n := \int_{\SpaceOfMotion^n} \mleft|\eta \mleft(x^{(n)}\mright)\mright| \,dx^{(n)} \,.
\end{equation}
\rev{For $n=0$, there is no dependence on the spatial position, so we have $\rho_0(t) \in \mathbb{R}$  and set $\|\rho_0(t)\|_0 := |\rho_0(t)|$.}

As the $n$ particles are of the same species and thus \rev{chemically} indistinguishable, we consider two-particle configurations as equivalent if they differ only by a permutation of particle indices. Restricting to this reduced configuration space, the relevant probability densities are symmetric \rev{under interchanging any pair of their arguments.
The same symmetry is known for bosonic many-body wave functions in quantum mechanics, albeit of totally different physical origin. The statistical indistinguishability of classical particles was introduced already by Gibbs to resolve the mixing paradox in statistical mechanics \cite{reif1965statphys}.
}

The open system with a variable number of particles is then characterized by the family of symmetrized probability densities,
\begin{equation} \label{eq:rhon_family}
	\rho(t)=(\rho_0(t), \rho_1(\cdot,t), \dots, \rho_n(\cdot,t), \dots) \,.
\end{equation}
Note that we will sometimes skip the time argument $t$ in the notation.
Recalling that $\|\rho_n(\cdot, t)\|_n = \Prob{N(t)=n}$, the total probability theorem implies that
\begin{equation} \label{eq:cons_prob}
	\sum_{n=0}^\infty \|\rho_n(\cdot, t)\|_n = 1 \quad \text{for all $t \geq 0$,}
\end{equation}
which expresses the conservation of probability also within chemical reaction.
The underlying space of such a family of distributions has a specific algebraic structure and is called the \textit{Fock space}; it will be introduced in detail in \cref{sec:fockspace}.
\rev{A similar probabilistic description, albeit differing in detail, was employed recently in a model of open systems that are coupled explicitly to a spatially separated reservoir \cite{dellesite2020liouville,klein2021nonequilibrium}.
In particular, the many-particle densities therein obey a set of coupled Liouville-type evolution equations that preserve total probability and the particle exchange symmetry.
}

We emphasize that the family of densities, \cref{eq:rhon_family}, must not be confused with the hierarchy of correlation functions $f_1, f_2, \dots$ that is used in statistical mechanics \rev{and is governed by the BBGKY equations} \cite{Hansen:SimpleLiquids}, where $f_n$ is also referred to as the reduced $n$-particle (phase space) density. Specifically, $f_1$ is the probability density of a tagged particle's position and $f_2$ refers to the correlation between a pair of particles \emph{in the presence of all other particles}.
These functions are obtained by marginalization of the distribution of the \rev{overall system} that consists of a large, but fixed number of particles.
In particular, $f_1$ can be obtained from $f_2$ by integration, which is conceptually different from the present approach, where $\rho_1$ and $\rho_2$ do not share such a relation.

\subsection{Chemical diffusion master equation}
\label{sec:intuitiveCDME}

The main goal of this work is to develop a \rev{framework to formulate} the evolution equation for the family of symmetrized $n$-particle densities in the presence of diffusion and reactions.

In the absence of reactions, the particle number is fixed and the system of interest may be associated with a closed system of $n$ particles that diffuse in space, possibly subject to physical interactions or an imposed flow.
In this case, the conditional density $p_n(x^{(n)},t|N(t)=n)$ of $n$-particle configurations $x^{(n)} \in \mathbb{X}^n$ obeys the Fokker-Planck equation
\begin{equation}
	\frac{\partial p_n}{\partial t} = \mathcal{D}_n p_n,
	\label{fokkplanck}
\end{equation}
where $\mathcal{D}_n$ is the corresponding Fokker-Planck operator. In the most general situation, it is a semi-elliptic linear operator and takes the form \cite{dhont1996dynamics}
\begin{equation}
	\mathcal{D}_n p_n = -\sum_{i=1}^{n} \nabla_i \cdot \left(A_i p_n\right) + \sum_{i,j=1}^{n} \nabla_i \cdot \left(D_{ij} \nabla_j p_n \right),
\end{equation}
where $A_i=A_i(x^{(n)},t)$ describes the deterministic drift, the $D_{ij} = D_{ij}(x^{(n)})$ are $3\times 3$ diffusion matrices composing the $3n \times 3n$ diffusion tensor, and $\nabla_i$ denotes differentiation with respect to the position $x_i^{(n)}$ of particle $i$. If the drift originates from an interaction potential $U(x^{(n)})$, then
\begin{equation}
	A_i = - \sum_{j=1}^{n} D_{ij} \nabla_j U.
\end{equation}
In the absence of the drift term and assuming that diffusion is isotropic, the diffusion operator reduces to that of standard Brownian motion,
$\mathcal{D}_n \rho_n = D\Delta \rho_n$,
with a scalar diffusion constant $D > 0$, where $\Delta$ is the Laplace operator.

\begin{figure*}
	\includegraphics[width=.8\textwidth]{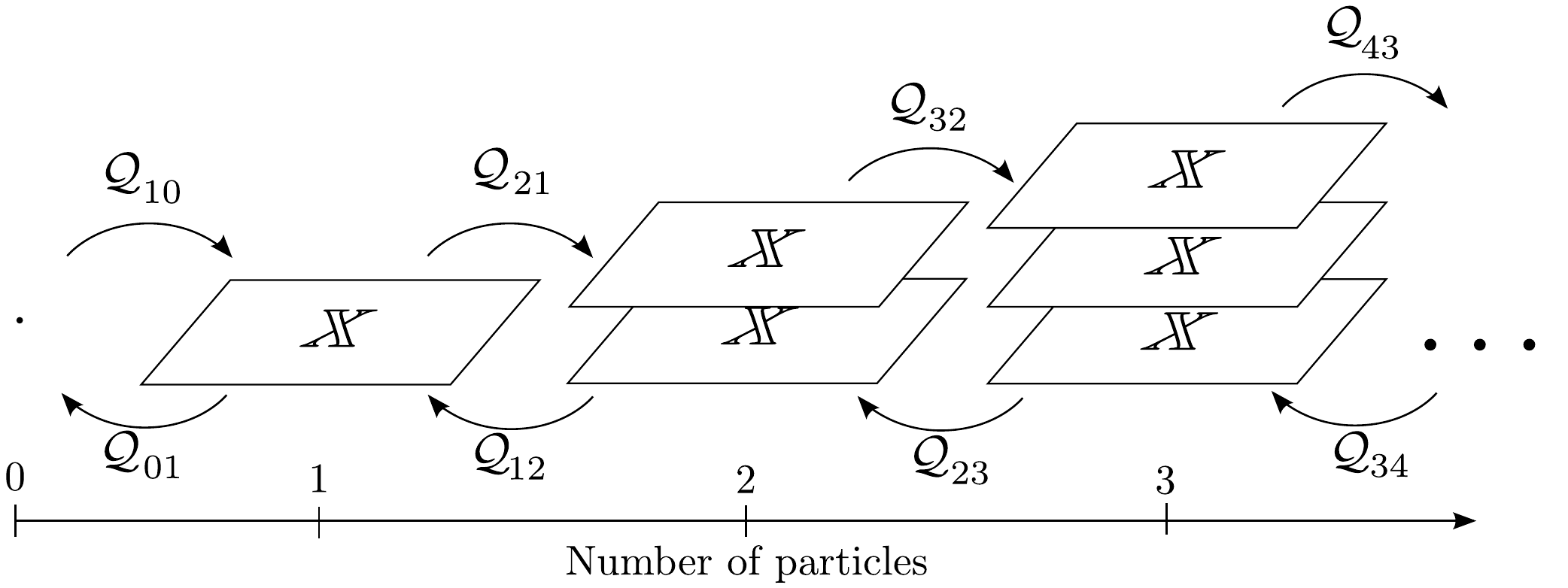}
	\caption{Illustration of the \rev{configuration} space of the stochastic process represented by the chemical diffusion-master equation \eqref{eq:mainMasterEq}. The \rev{position} space for every particle is $\SpaceOfMotion \subset \mathbb{R}^3$. The \rev{configuration} space is composed of subspaces $\SpaceOfMotion^{n}$ of dimension $n=0,1,2,\dots$, which is equal to the number of particles in the system.
		Chemical reactions lead to transitions between these subspaces by adding or removing particles with position-dependent rates as encoded in the matrix $(\AltReactionOperator_{nm})$ of reaction operators.
		$\AltReactionOperator_{nm}$ represents the transition from a \rev{configuration} with $m$ particles to a \rev{configuration} with $n$ particles.
		For simplicity, only transitions that differ by one particle are shown.
	}
	\label{fig:diagPhaseSp}
\end{figure*}

\rev{In addition to diffusive motion, the particles of the considered species $\Species$ undergo reactions of the form
  \begin{equation}
	k\Species\rightarrow l \Species, \quad k,l\in \mathbb{N}_0.
  \end{equation}
  \rrev{For each reaction, there is a \textit{reaction rate function} 
  	\begin{equation}\label{propensity}
  	\Propensity:\SpaceOfMotion^l\times \SpaceOfMotion^k \to [0,\infty),
  \end{equation}
	where the value $\Propensity(y^{(l)},x^{(k)})\geq 0$ corresponds to the rate at which the reaction} occurs given that the $k$ reactants are at positions $x^{(k)}\in \SpaceOfMotion^k$ and the $l$ products are placed at positions $y^{(l)}\in \SpaceOfMotion^l$. The reaction rate function $\Propensity$ is assumed to be symmetric under particle exchange within the configuration $y^{(l)}$ and also within $x^{(k)}$.}

In the presence of both reactions and diffusion, the total number of particles and their positions can change simultaneously over time. 
Given that there are $S$ different chemical reactions, the family of $n$-particle densities, $\rho = (\rho_0, \rho_1, \dots)$, should obey a linear evolution equation of the form
\begin{align}\label{eq:emptyCDME}
\partial_t \rho = \left(\mathcal{D}  +  \sum_{r=1}^S \ReactionOperator_r\right) \rho
\end{align}
which we will refer to as the \emph{chemical diffusion master equation} (CDME).
Here, $\mathcal{D}$ is the diffusion operator as above, and the reaction operator $\ReactionOperator_r$ encodes the $r$-th reaction, which is parametrized by a reaction rate function $\lambda_r$.
A central assumption is that diffusion and reactions occur independently, which allows us to split the operator on the right-hand side into the sum $\mathcal{D} + \ReactionOperator$,
where $\ReactionOperator = \sum_{r=1}^S \ReactionOperator_r$ combines all reaction operators into one to simplify the notation.
The reaction operators couple components $\rho_n$ of $\rho$ for different $n$, similarly to the CME.
More precisely, treating $\rho$ as an infinitely long column vector, \cref{eq:emptyCDME} reads in matrix notation:
\rev{
\begin{align}
	\frac{\partial}{\partial t} 
	\underbrace{
		\begin{pmatrix}
			\rho_0 \\
			\rho_1 \\
			\vdots \\
			\rho_n \\
			\vdots
	\end{pmatrix}}_{\textstyle\rho}
	=
	\underbrace{ 
		\begin{pmatrix}
			\mathcal{D}_0\rho_0 \\
			\mathcal{D}_1\rho_1 \\
			\vdots \\
			\mathcal{D}_n\rho_n \\
			\vdots
	\end{pmatrix}}_{\textstyle\mathcal{\mathcal{D}}\rho}
	+
	\underbrace{\begin{pmatrix}
			\AltReactionOperator_{00} & \AltReactionOperator_{01} & \dots & \AltReactionOperator_{0n} & \dots\\
			\AltReactionOperator_{10} & \AltReactionOperator_{11} & \dots & \AltReactionOperator_{1n} & \dots \\
			\vdots  & \vdots && \vdots& \dots \\
			\AltReactionOperator_{n0} & \AltReactionOperator_{n1} & \dots & \AltReactionOperator_{nn} & \dots\\
			\vdots & \vdots & &\vdots & \ddots
	\end{pmatrix}}_{\textstyle \ReactionOperator}
	\begin{pmatrix}
		\rho_0 \\
		\rho_1 \\
		\vdots \\
		\rho_n \\
		\vdots
	\end{pmatrix}.
	\label{eq:mainMasterEq}
\end{align}
Each entry $\AltReactionOperator_{nm}$ is an operator that condenses the effect of all the reactions that act on a given $\rho_{n'}$ yielding the temporal change of $\rho_n$, similarly to the CME.} 
To ensure conservation of probability [\cref{eq:cons_prob}] we impose reflective boundary conditions at the boundaries of the domains $\SpaceOfMotion^n$ separately for every $n$,
i.e., the diffusive flux across these boundaries is zero.
Under the assumption of well-mixed conditions, the spatial degrees of freedom can be integrated out, and \cref{eq:mainMasterEq} should yield the conventional CME \cite{vanKampen1992stochastic}. On the other hand, if there are no reactions, the equations will uncouple and yield a set of independent Fokker-Planck equations for the symmetrized densities $\rho_n$, see \cref{fokkplanck}.

The stochastic system trajectories corresponding to \cref{eq:mainMasterEq} combine continuous diffusion in the \rev{configuration} spaces $(\mathbb{X}^0, \mathbb{X}^1, \dots, \mathbb{X}^n, \dots)$ for fixed $n$ and a discrete jump process between $\mathbb{X}^m$ and $\mathbb{X}^n$ for $m\neq n$, modelling creation and annihilation of particles due to reactions (\cref{fig:diagPhaseSp}).
The model thus fits into the category of stochastic hybrid systems, where continuous dynamics and discrete events coexist in the same process. However, a particular challenge for the mathematical formalization is the change of dimensionality that is induced by the discrete jumps.

\paragraph*{Example (birth-death process).}
Consider a chemical species $\Species$ that undergoes degradation and creation reactions:
\begin{equation} \label{eq:degradation_creation_reactions}
	\mathrm{(I)} \quad \Species \xrightarrow{\:\lambda_d(x)\:} \emptyset, \qquad
	\mathrm{(II)} \quad \emptyset \xrightarrow{\:\lambda_c(x)\:} \Species   \,.
\end{equation}
Here, $\Propensity_d(x)$ denotes the rate for reaction $\mathrm{(I)}$ to occur for a particle located at position $x\in \SpaceOfMotion$ (i.e., the probability per unit of time for this particle to disappear), while $\Propensity_c(x)$ is the rate for a new particle to be created at position $x$ by reaction $\mathrm{(II)}$.
Explicitly, we assume that the rates depend only on the position in space, but not on the configuration of all particle positions. \rrev{Both $\Propensity_d$ and $\Propensity_c$ are special cases of the generic reaction rate function $\Propensity$ in \eqref{propensity}.}

\rrev{Arguments analogous to the formulation of the CME lead us to propose} the following equation for the time evolution of $\rho_n(x^{(n)},t)$ for $n\geq 1$, \rrev{taking into account the symmetry of the densities:}
\begin{align}\begin{split} \label{eq:time_evolution}
		\partial_t \rho_n(x^{(n)}_1,\dots,x^{(n)}_n,t) &=
		\mathcal{D}_n\rho_n(x^{(n)}_1,\dots,x^{(n)}_n,t) \\
		& \quad + (n+1)\int_\SpaceOfMotion \Propensity_d(y)  \rho_{n+1}(x^{(n)}_1,\dots,x^{(n)}_n,y,t) \, dy  \\
		& \quad  -\sum_{i=1}^n \Propensity_d(x^{(n)}_i)  \rho_n(x^{(n)}_1,\dots,x^{(n)}_n,t)   \\
		& \quad + \frac{1}{n}\sum_{i=1}^n \Propensity_c(x^{(n)}_i)\rho_{n-1}(x^{(n)}_1,\dots,x^{(n)}_{i-1},x^{(n)}_{i+1},\dots,x^{(n)}_n,t)  \\
		& \quad - \int_\SpaceOfMotion \Propensity_c(y)  \rho_n(x^{(n)}_1,\dots,x^{(n)}_n,t) \, dy \,,
\end{split}\end{align}
where the first line refers to spatial diffusion, the second and third lines are gain and loss terms due to reaction $\mathrm{(I)}$, and the last two lines relate to reaction $\mathrm{(II)}$.
For $n=0$, we have
\begin{equation} \label{eq:time_evolution0}
	\partial_t \rho_0(t) = \int_{\SpaceOfMotion} \Propensity_d(x) \rho_1(x,t) dx  -\int_\SpaceOfMotion \Propensity_c(x) \rho_0(t) \, dx.
\end{equation}
The boundary-value problem induced by \cref{eq:time_evolution,eq:time_evolution0} is well posed for an elliptic diffusion operator $\mathcal{D}$ and regular reaction rate functions $\lambda_\mathrm{d}$ and $\lambda_\mathrm{c}$.
\Cref{eq:time_evolution} corresponds to the $n$-th row of \cref{eq:mainMasterEq}, and the components of the reaction operator matrix are:
\begin{subequations}
\begin{align}
	(\AltReactionOperator_{n,n+1}\rho_{n+1})(x^{(n)}) &= \quad (n+1)\int_\SpaceOfMotion \Propensity_d(y)  \rho_{n+1}(y,x^{(n)}_1,\dots,x^{(n)}_n) \, dy, \label{eq:Rbirthdeath1} \\
	(\AltReactionOperator_{n,n-1}\rho_{n-1})(x^{(n)}) &= \quad \frac{1}{n} \sum_{i=1}^n \Propensity_c(x^{(n)}_i)\rho_{n-1}(x^{(n)}_1,\dots,x^{(n)}_{i-1},x^{(n)}_{i+1},\dots,x^{(n)}_n), \label{eq:Rbirthdeath2} \\
	(\AltReactionOperator_{n,n}\rho_n)(x^{(n)}) &= - \left [\sum_{i=1}^n \Propensity_d(x^{(n)}_i)   +   \int_\SpaceOfMotion \Propensity_c(x) \, dx \right] \rho_n(x^{(n)}_1,\dots,x^{(n)}_n),\label{eq:Rbirthdeath3}
\end{align}
\end{subequations}
and for $n=0$, they are
\begin{subequations}
\begin{align}
	\AltReactionOperator_{0,1}\rho_1 &= \quad \int_{\SpaceOfMotion} \Propensity_d(x) \rho_1(x) dx , \\
	\AltReactionOperator_{0,0}\rho_0 &= -\int_\SpaceOfMotion \Propensity_c(x)\rho_0 \, dx ,
\end{align}
\end{subequations}
where $\rho_0\in \mathbb{R}$ is a constant.

We verify that the CME related the reactions (I) and (II) is recovered from \cref{eq:time_evolution} by integrating out the spatial degrees of freedom [\cref{P(N)}]. To this end, we define the marginal distribution
\begin{equation}\label{def:marginaldistr}
	P(n,t):= \Prob{N(t)=n}= \| \rho_n(\cdot, t) \|_n
\end{equation}
and assume the reaction rate functions to be constants,
$\Propensity_d(x)=\gamma_\mathrm{d}$ and $\Propensity_c(x)=\gamma_\mathrm{c}$ for all $x\in\SpaceOfMotion$.
Then integrating \cref{eq:time_evolution} over the space $\SpaceOfMotion^n$ yields:
\begin{align} \label{eq:example_cme}
	\partial_t P(n,t) &= (n+1)\gamma_\mathrm{d} P(n+1,t) - n\gamma_\mathrm{d} P(n,t) \nonumber \\
	& \quad + \gamma_\mathrm{c} |\SpaceOfMotion| P(n-1,t) - \gamma_\mathrm{c} |\SpaceOfMotion| P(n,t),
\end{align}
where $|\SpaceOfMotion|<\infty$ is the volume of the domain $\SpaceOfMotion$.
Note that the diffusion term vanishes due to the no-flux boundary condition by Gauss' theorem.
\Cref{eq:example_cme} is exactly the CME as derived from the classical law of mass action
for reactions of order zero and one in spatially well-mixed systems.
One checks easily that summation of the right-hand side over $n \geq 0$ yields zero, as required for a continuous-time Markov chain.

In the next two sections, we will develop a systematic way to construct the CDME such as \cref{eq:time_evolution} corresponding to a given set of chemical reactions [e.g., \cref{eq:degradation_creation_reactions}].
In particular, we will specify the reaction operators [\cref{eq:Rbirthdeath1,eq:Rbirthdeath2,eq:Rbirthdeath3}] and find their correct combinatorial prefactors, which is often not a straightforward task. \rev{We do so by means of the Fock space formalism, which yields one possible explicit representation of the probabilistic evolution described by the CDME. }

\section{Fock space formalism}
\label{sec:fockspace}

The mathematical formalization of the dynamics of open systems requires means for the insertion and deletion of particles. In reaction-diffusion problems, this can occur everywhere in the domain $\mathbb{X}$. Further, in the probabilistic description, a particle is described by a probability density rather than by a single point.
A similar problem was solved in quantum field theory, where the $n$-particle \rev{densities} are
represented by symmetric or anti-symmetric wavefunctions and (quasi-)particles can be created in or annihilated from such states.
The underlying algebraic construction is called \emph{Fock space} and
relies on the fact that the space of $n$-particle densities is generated from products of single-particle \rev{densities}.
In the quantum case, the wavefunctions are square-integrable and form a Hilbert space whereas the probability densities of classical particles are absolutely integrable [\cref{P(N)}]. As an important consequence, the corresponding space $L^1(\mathbb{X}^n)$ is merely a Banach space, i.e., it does not possess an inner product.
This technical deficiency can be circumvented by resorting to the dual space.
\rev{In the following, we will introduce the Fock space for the family of symmetrized probability densities. Suitably defined creation and annihilation operators will then serve as efficient instruments to represent the reaction operators in the CDME as well as its solutions. This program will be carried out in \cref{sec:rep_operators}.
}

\subsection{Symmetrized $n$-particle spaces}
\label{sec:symm-npart-spaces}

\paragraph*{Single-particle space.}

We start with the one-particle space $\FunctionSpace:=L^1(\SpaceOfMotion)$, which contains the probability density functions of a single particle's position on the space of motion $\SpaceOfMotion$,
and re-collect some facts from functional analysis \cite{werner2006funktionalanalysis}.
The dual space of $L^1(\SpaceOfMotion)$ is isometrically isomorph to $L^\infty(\SpaceOfMotion)$, so that we can identify the dual space $\FunctionSpace^*$ with $L^\infty(\SpaceOfMotion)$, i.e., the bounded functions on $\SpaceOfMotion$.
In the application to reaction-diffusion dynamics below, we will see that, for example, reaction rate functions are elements of $H^*$.
A dual pairing $\langle \cdot,\cdot\rangle: \FunctionSpace^* \times \FunctionSpace \to \mathbb{R}$ is defined by
\begin{equation}\label{dual_pairing_1}
	\langle \zeta, \eta\rangle := \int_{\SpaceOfMotion} \zeta(x) \eta(x) dx \,,
\end{equation}
which acts as a substitute for the missing inner product on $H$ and yields the ``overlap'' between the functions $\zeta \in \FunctionSpace^*$ and $\eta\in\FunctionSpace$.
The next step is to choose a (Schauder) basis $(u_1, u_2, \dots)$ of the space $H$, which can even be taken to be non-negative \cite{johnson2015schauder} and normalized, $u_\alpha \geq 0$ and $\|u_\alpha\|_1 = 1$ for all $\alpha \in \mathbb{N}$.
The basis induces the dual set of linear functionals $(u_1^*, u_2^*, \dots)$, here identified with functions $u^*_\alpha \in L^\infty(\SpaceOfMotion)$, such that $ \langle u^*_\alpha, u_\beta \rangle = \delta_{\alpha,\beta} $ for all $\alpha, \beta \in \mathbb{N}$, using the Kronecker symbol $\delta_{\alpha,\beta}$.
Thus, a density $\eta \in \FunctionSpace$ of a single particle has the representation
\begin{align}
	\eta = \sum_{\alpha=1}^\infty \langle u_\alpha^*, \eta\rangle u_\alpha
	\label{eq:expansionHelement}
\end{align}
in terms of the one-particle basis $(u_\alpha)_{\alpha\in\mathbb{N}}$.
We note that the family $(u_{\alpha}^*)_{\alpha\in\mathbb{N}}$ is countable and thus cannot span the dual space $H^* \cong L^\infty(X)$, which is not separable. However, this deficiency of $(u_\alpha^*)$ is not of practical relevance for the following treatment.

\paragraph*{Tensor spaces.}

A natural extension to the space $L^1(\mathbb{X}^n)$ of $n$-particle densities uses the observation that
the \textit{pure tensor} $u_{\alpha_1} \otimes \dots \otimes u_{\alpha_n}$, given as products of $n$ one-particle densities,
\begin{equation}
	(u_{\alpha_1} \otimes \dots \otimes u_{\alpha_n})(x^{(n)}) := u_{\alpha_1}(x^{(n)}_1) \ldots u_{\alpha_n}(x^{(n)}_n) \,,
\end{equation}
span the space $L^1(\mathbb{X}^n)$. Thus, the tensor space
\begin{equation}
	\FunctionSpace^{\otimes n} := \bigotimes_{i=1}^n\FunctionSpace
	= \overline{\Span} (u_{\alpha_1} \otimes \dots \otimes u_{\alpha_n})_{\alpha_i \in \mathbb{N}}
\end{equation}
coincides with $L^1(\mathbb{X}^n)$ and we refer to $(u_{\alpha_1} \otimes \dots \otimes u_{\alpha_n})_{\alpha_i \in \mathbb{N}}$ as a tensor basis.
As convention for $n=0$, we set $\FunctionSpace^{\otimes 0} := \mathbb{R}$.
Similarly to the one-particle case, we define a dual pairing for $\eta\in \FunctionSpace^{\otimes n}$, $\zeta\in (\FunctionSpace^{\otimes n})^*$ as
\begin{equation} \label{eq:altInnerProd2}
	\langle \zeta, \eta \rangle :=  \int_{\SpaceOfMotion^n}  \zeta(x^{(n)}) \, \eta(x^{(n)}) \, d x^{(n)}.
\end{equation}
Then the dual set of the tensor basis consists of the dual pure tensor
\begin{equation}
	(u_{\alpha_1} \otimes \dots \otimes u_{\alpha_n})^* = u_{\alpha_1}^* \otimes \dots \otimes u_{\alpha_n}^* \,,
\end{equation}
which satisfy
\begin{align} \label{eq:bi-orthogonality}
	\langle u_{\alpha_1}^* \otimes \dots \otimes u_{\alpha_n}^*, u_{\alpha_1'} \otimes \dots \otimes u_{\alpha_n'} \rangle
	& = \langle u_{\alpha_1}^*, u_{\alpha_1'} \rangle \dots \langle u_{\alpha_n}^*, u_{\alpha_n'} \rangle \nonumber \\
	& = \delta_{\alpha_1 \alpha_1'} \dots \delta_{\alpha_n \alpha_n'}
\end{align}
for all multi-indices $(\alpha_1, \dots, \alpha_n), (\alpha_1', \dots, \alpha_n') \in \mathbb{N}^n$.
Finally, the basis representation of an element $\eta \in \FunctionSpace^{\otimes n}$ reads
\begin{equation}\label{eq:v_basis2}
	\eta=\sum_{\alpha_1, \dots , \alpha_n} c_{\alpha_1,\dots,\alpha_n} \, u_{\alpha_1} \otimes \dots \otimes u_{\alpha_n} \,,
\end{equation}
with coefficients
$c_{\alpha_1,\dots,\alpha_n} := \langle u_{\alpha_1}^* \otimes \dots \otimes u_{\alpha_n}^*,\eta \rangle$
and each of the summation indices $\alpha_1 \dots \alpha_n$ running from 1 to $\infty$.

\paragraph*{Symmetrization.}

\rev{The probability density of a system of $n$ identical particles is symmetric in all particle positions (see  \cref{sec:probdens}). Thus, the spaces $H^{\otimes n}$ are too large for our purposes and we need to project onto the symmetrized functions. To this end, we introduce the symmetrization operator $\SymmOperator_n:  \FunctionSpace^{\otimes n} \rightarrow  \FunctionSpace^{\otimes n}$,
\begin{equation}\label{Symm}
	\SymmOperator_n \eta\mleft(x^{(n)}_1,\dots,x^{(n)}_n,t\mright) \\
	:= \frac{1}{n!}\sum_{\sigma \in \Sigma_n}  \eta\mleft (x^{(n)}_{\sigma(1)}, \dots, x^{(n)}_{\sigma(n)},t\mright ),
\end{equation}
where $\Sigma_n$ is the set of permutations on $\{1,\dots,n\}$. We note that for $n=0$ or $n=1$, the symmetrization is the identity: $\SymmOperator_0\eta = \eta$ for $\eta \in \FunctionSpace^{\otimes 0}$ or $\eta \in \FunctionSpace^{\otimes 1}$.
Dividing by the number $n!$ of possible permutations ensures that normalization is preserved, i.e.,
$\|\SymmOperator_n \eta\|_n = \|\eta\|_n$.  
The operator $\SymmOperator_n$ is indeed a linear projection, in particular $\SymmOperator_{n}^2 = \SymmOperator_{n}$,
and it is also ``orthogonal'' with respect to the dual pairing (proof in \cref{app:dualpairingsymmetric}):
\begin{equation} \label{eq:symm-orth}
\langle \SymmOperator_{n} \zeta, \eta \rangle = \langle \zeta, \SymmOperator_{n} \eta \rangle
= \langle \SymmOperator_{n} \zeta, \SymmOperator_{n} \eta \rangle
\end{equation}
for any $\eta \in \FunctionSpace^{\otimes n}$ and $\zeta \in (\FunctionSpace^{\otimes n})^*$. Throughout this work we will refer to the symmetrized tensor space $\SymmOperator_n\FunctionSpace^{\otimes n}= \{ \SymmOperator_n v : v \in \FunctionSpace^{\otimes n} \}$ as the \emph{$n$-particle space.}
}

A basis of $\SymmOperator_n\FunctionSpace^{\otimes n}$ is obtained by symmetrization of the $n$-particle tensor basis:
\begin{equation}
	\bigl(\SymmOperator_n (u_{\alpha_1} \otimes \dots \otimes u_{\alpha_n}) \bigr)_{\alpha_1\leq \dots \leq \alpha_n},
\end{equation}
where the action of $\SymmOperator_n$ on a pure tensor amounts to a linear combination of all permutations of the factors $u_{\alpha_1}, \dots, u_{\alpha_n}$.
The ordering of the indices $\alpha_1 \leq \dots \leq \alpha_n$ is needed to avoid double-counting;
the equality is included here as, for example, $u_1 \otimes u_1$ belongs to $\SymmOperator_2 H^{\otimes 2}$.
The dual set of the symmetrized tensor basis obeys again a bi-orthogonality relation [see \cref{eq:bi-orthogonality}]:
\begin{align}
	\langle
	\SymmOperator_n( u_{\alpha_1} \otimes \dots \otimes u_{\alpha_n})^*,
	\SymmOperator_n(u_{\alpha_1'} \otimes \dots \otimes u_{\alpha_n'})
	\rangle = \delta_{\alpha_1 \alpha_1'} \dots \delta_{\alpha_n \alpha_n'}
\end{align}
with the understanding that in the expression $\SymmOperator_n( \dots )^*$ dualization is carried out \emph{after} symmetrization.
It is evident from \cref{eq:symm-orth} that the two operations do not commute, which motivates us to introduce ``normalization'' coefficients
\begin{equation} \label{eq:symm-coeff}
	s_{\alpha_1,\dots,\alpha_n} := \langle
	\SymmOperator_n( u_{\alpha_1}^* \otimes \dots \otimes u_{\alpha_n}^*),
	\SymmOperator_n(u_{\alpha_1} \otimes \dots \otimes u_{\alpha_n})
	\rangle \,,
\end{equation}
so that the symmetrized duals are given in terms of the unsymmetrized duals by
\begin{equation} \label{eq:symm-duals}
	\SymmOperator_n( u_{\alpha_1} \otimes \dots \otimes u_{\alpha_n})^*
	= \frac{1}{s_{\alpha_1,\dots,\alpha_n} } \SymmOperator_n( u_{\alpha_1}^* \otimes \dots \otimes u_{\alpha_n}^*) \,.
\end{equation}
As an example, for $n=2$, one calculates
\begin{align}
	s_{\alpha\beta} &= \langle \SymmOperator_{2} (u_\alpha^* \otimes u_\beta^*), \SymmOperator_{2} (u_\alpha \otimes u_\beta) \rangle \nonumber \\
	&= \left\langle \frac{1}{2}(u_\alpha^* \otimes u_\beta^* + u_\beta^* \otimes u_\alpha^*),
	\frac{1}{2}(u_\alpha \otimes u_\beta + u_\beta \otimes u_\alpha) \right\rangle \nonumber \\
	&= \frac{1}{2}(1 + 1\delta_{\alpha\beta}) \label{s_alphabeta}
\end{align}
so that $s_{\alpha\beta} = \frac{1}{2}$ for $\alpha\neq\beta$ and $s_{\alpha\beta} = 1$ otherwise.
With this, a symmetrized function $\eta \in \SymmOperator_n\FunctionSpace^{\otimes n}$ has the basis representation
\begin{equation}\label{basis_representation}
	\eta=\sum_{\alpha_1\leq\dots\leq \alpha_n}\tilde c_{\alpha_1,\dots,\alpha_n} \SymmOperator_n (u_{\alpha_1} \otimes \dots \otimes u_{\alpha_n})
\end{equation}
with coefficients
\begin{align}
	\tilde c_{\alpha_1,\dots,\alpha_n}
	&:= \langle \SymmOperator_n (u_{\alpha_1} \otimes \dots \otimes u_{\alpha_n})^*, \eta \rangle \nonumber \\
	&~= \frac{1}{s_{\alpha_1,\dots,\alpha_n} } \langle u_{\alpha_1}^* \otimes \dots \otimes u_{\alpha_n}^*, \eta \rangle
\end{align}
for $1\leq \alpha_1\leq\dots\leq \alpha_n < \infty$, using \cref{eq:symm-duals,eq:symm-orth} in the second line.
For later reference, we again define the subspace spanned by the duals of the symmetrized tensor basis:
\begin{equation}
	(\SymmOperator_n \FunctionSpace^{\otimes n})'
	:= \overline{\Span}\bigl(\SymmOperator_n (u_{\alpha_1} \otimes \dots \otimes u_{\alpha_n})^*\bigr)_{\alpha_i \in \mathbb{N}}
	\subset (\SymmOperator_n H^{\otimes n})^* \,.
\end{equation}

\paragraph*{Copy number representation.} \label{copy_number}

In the basis representation \eqref{basis_representation}, the indices $\alpha_1,\dots,\alpha_n$ \rev{take values that are not necessarily different from each other.} This motivates counting the occurrence of every value $\alpha=1, 2, \dots$ in a given multi-index $(\alpha_1,\dots,\alpha_n)$ and switching to a sequence of frequencies $(N_1, N_2, \dots)$, where most of the entries are zero.
We refer to these frequencies as \emph{copy numbers} since $N_\alpha$ counts how many times the factor $u_\alpha$ is repeated in an element of the tensor basis, i.e., how many particles \rev{have a position distributed according to $u_\alpha$}; 
in quantum mechanics, the term occupation number is used instead.
Given the number of particles $n\in \mathbb{N}$, we define
\begin{equation}\label{M_n}
	\mathbb{M}_n := \left\{N=(N_1,N_2,\dots): N_\alpha \in \mathbb{N}_0, \sum_{\alpha=1}^{\infty} N_\alpha =n \right\}
\end{equation}
as the set of possible sequences of copy numbers and use it to enumerate the symmetrized $n$-particle tensor basis.
Then, for $\eta \in \SymmOperator_n\FunctionSpace^{\otimes n}$, we have
\begin{equation}
	\eta = \sum_{N\in\mathbb{M}_n }  p_{N} \,
	\SymmOperator_n(u_1^{\otimes N_1}\otimes  u_2^{\otimes N_2}\otimes\dots)
	\label{eq:copynumBasisExp}
\end{equation}
with coefficients
$p_{N} = \bigl\langle \SymmOperator_n(u_1^{\otimes N_1}\otimes  u_2^{\otimes N_2}\otimes\dots)^*, \eta \bigr\rangle$ .
Expanding the symmetrization operator and contracting the dual pairings in \cref{eq:symm-coeff}, one can show that
\begin{equation}
	\SymmOperator_n (u_1^{*\otimes N_1}\otimes  u_2^{*\otimes N_2}\otimes\dots) = \frac{N_1! \dots N_M!}{n!} \,
	\SymmOperator_n(u_1^{\otimes N_1}\otimes  u_2^{\otimes N_2}\otimes\dots)^*
\end{equation}
so that
\begin{equation}
	p_{N} = \frac{n!}{N_1!N_2! \dots } \, \left\langle u_1^{*\otimes N_1}\otimes  u_2^{*\otimes N_2}\otimes\dots, \eta\right \rangle ,
\end{equation}
where $\SymmOperator_n$ has been omitted in the left factor on the r.h.s.\ since $\eta$ is symmetrized.
As $N_\alpha$ is non-zero for at most $n$ values of $\alpha$ (and recalling that $0! = 1$ in all other cases), the multi-nomial coefficients $n!/(N_1!N_2! \dots)$ are well defined. Note these multinomial coefficients simply correspond to the ``normalization''coefficients defined in \cref{eq:symm-coeff}.

\subsection{The Fock space}

The \emph{Fock space} $\Fockspace$ holds the probability densities on the configuration space of the open system, depicted in \cref{fig:diagPhaseSp}.
\rrev{The elements of $\Fockspace$ are families $\rho=(\rho_0,\rho_1,\rho_2,\dots)$ of symmetrized $n$-particle densities, which form}
the direct sum of symmetrized $n$-particle spaces (see \cref{sec:symm-npart-spaces}),
\begin{equation} \label{def:fockspace}
  \bigoplus_{n=0}^\infty \SymmOperator_n \FunctionSpace^{\otimes n} = \left\{
    \rho=(\rho_0,\rho_1,\rho_2,\dots) :
      \rho_n\in  \SymmOperator_n \FunctionSpace^{\otimes n} \mbox{ for all } n\in \mathbb{N}_0
  \right \} .
\end{equation}
It is sometimes convenient to interpret $\rho_n \in  \SymmOperator_n \FunctionSpace^{\otimes n}$ as the Fock space element $(0,\dots,0,\rho_n,0,\dots)$ and to introduce a componentwise addition on $\Fockspace$, which allows for a compact notation such as in
\begin{equation}
\rho= \sum_{n=0}^\infty \rho_n \,.
\label{sum_notation}
\end{equation}
\rrev{Further, we introduce the following generalization of the $L^1$-norm,
\begin{equation} \label{norm}
	\|\rho\| :=  \sum_{n=0}^\infty \|\rho_n \|_n, \,
\end{equation}
with $\|\cdot \|_n$ as in \cref{n-norm},
and define $\Fockspace$ as the set of probability densities of the open system that are integrable and absolutely summable:
\begin{equation}
  \Fockspace := \left\{\rho \in \bigoplus\nolimits_{n=0}^\infty \SymmOperator_n \FunctionSpace^{\otimes n} : \|\rho\| <\infty \right\}.
\end{equation}
}%
In reaction-diffusion problems, we will consider probability densities $\rho \in\Fockspace$ which are normalized,
$\|\rho\| = 1$, and preserve this normalization under time evolution [see \cref{eq:cons_prob}].

In contrast to the $L^2$-norm used in quantum mechanics, \rrev{the $L^1$-norm on $H$ and likewise the norm $\|\cdot \|$ on $\Fockspace$} is not induced by any inner product. \rrev{Hence, $\Fockspace$ cannot be a Hilbert space, but is merely a Banach space.}
The dual space of $\Fockspace$ is given by
\begin{equation}
  \Fockspace^* := \left\{ \nu \in \bigoplus\nolimits_{n=0}^\infty \SymmOperator_n (\FunctionSpace^{\otimes n})^* :
    \rrev{\|\nu\|_\infty < \infty} \right\}
\end{equation}
with elements $\nu=(\nu_0, \nu_1,\nu_2,\dots)$ \rrev{and the supremums norm,
\begin{equation}
  \|\nu\|_\infty := \sup_n \sup_{x\in \SpaceOfMotion^n} |\nu_n(x)| \,.
\end{equation}
}%
The dual pairing for $\rho \in \Fockspace$ and $\nu\in \Fockspace^*$ is
\begin{equation}\label{Scalarproduct1}
\langle \nu, \rho\rangle := \sum_{n=0}^\infty  \langle \nu_n, \rho_n\rangle,
\end{equation}
with $\langle \nu_n, \rho_n\rangle$ defined in \cref{eq:altInnerProd2}.

\subsection{Creation and annihilation operators} \label{sec:creat_annihil}

The creation and annihilation operators raise or lower the number of particles \rev{in the system by mapping an $n$-particle density to an $(n+1)$-particle density or $(n-1)$-particle density, respectively.}
The creation operator $a^+\{w\}$ adds a particle whose position is described by the probability density $w(x)\in  H$, and the annihilation operator $a^-\{f\}$ removes a particle with reaction rate function $f(x)\in H^*$.
These operators are defined on the symmetrized $n$-particle spaces and, most importantly, preserve symmetrization:
\begin{subequations}
\begin{align}
  a^+\{w\} &: \SymmOperator_n\FunctionSpace^{\otimes n} \to \SymmOperator_{n+ 1}\FunctionSpace^{\otimes (n+1)} , \\
  a^-\{f\} &: \SymmOperator_n\FunctionSpace^{\otimes n} \to \SymmOperator_{n- 1}\FunctionSpace^{\otimes (n-1)} .
\end{align}
\end{subequations}
There action on a symmetrized pure tensor
$v=\SymmOperator_n (v_1 \otimes \dots \otimes v_n) $ for $v_i \in H$
is defined as
\begin{subequations}
	\label[eqset]{def:longsymaj+-}
	\begin{align}
		a^+\{w\}v & := \SymmOperator_{n+1} (w\otimes v_1 \otimes \dots \otimes v_n) \,, \label{def:longsymaj+} \\
		a^-\{f\}v & := \sum_{j=1}^n \langle f,v_{j} \rangle \SymmOperator_{n-1}\left(v_{\setminus \{j\}}\right) , \label{def:longsymaj-}
	\end{align}
\end{subequations}
where $v_j$ is omitted in $v_{\setminus \{j\}}:=v_1\otimes \dots \otimes v_{j-1}\otimes v_{j+1}\otimes  \dots \otimes v_n$;
for $n=0$, we set $a^-\{f\} 1 := 0$.
By linearity, these definitions extend naturally to the space $\SymmOperator_n\FunctionSpace^{\otimes n}$, since the elements of its tensor basis are of the form $v$, and to the whole Fock space $\Fockspace$ by acting componentwise on $\rho$.

The sum in the annihilation operator [\cref{def:longsymaj-}] expresses the fact that there are $n$ different ways to remove a particle.
It reduces to a factor $n$ in the well-mixed case without spatial resolution (where all $v_i$ are equal to the uniform distribution).
The above definitions thus differ from the ones used for quantum systems, where $\sqrt{n+1}$ and $\sqrt{n}$ are the coefficients of $a^+$ and $a^-$, respectively. The present choice is suitable for systems of indistinguishable classical particles \cite{baez2018quantum,doi1976second,grassberger1980fock}, and for chemical systems, they avoid cumbersome prefactors in the subsequent results.

The creation and annihilation operators have expansions in the basis $(u_\alpha)$ of the single-particle space $H$ [see \cref{eq:aps_expansion_app,eq:ams_expansion_app}]:
\begin{subequations}
\begin{align}
 a^+\{w\} &= \sum_{\alpha} \langle u_\alpha^*, w \rangle a^+\{ u_\alpha \} \,, \label{eq:aps_expansion} \\
 a^-\{f\} &= \sum_{\alpha} \langle f, u_\alpha \rangle a^-\{u_\alpha^*\} \,. \label{eq:ams_expansion}
\end{align}
\end{subequations}
\rev{A basis-free representation of the action of the two operators on a symmetrized $n$-particle density is given in \cref{def:densityaj+2,def:densityaj-2}.}
We note that for a given annihilation operator $a^-\{f\}$ the choice of the reaction rate function $f$ is not unique. It is seen from
the second relation that $a^-\{f\} = a^-\{f + \zeta\}$ for any $\zeta \in \FunctionSpace^*$ such that $\langle \zeta, u_\alpha \rangle = 0$ for all $\alpha$. Such $\zeta \neq 0$ exist since $(u_\alpha^*)$ does not span $\FunctionSpace^*$.
However, this mathematical issue has no further implications for the present work.

\paragraph*{Operator algebra.}
The creation and annihilation operators satisfy a number of basic relations that define an operator algebra, which is useful to derive representations in terms of these operators.
\begin{enumerate}
	\item Removing a particle from the (normalised) vacuum \rev{element} $\rhovac := (1, 0, 0, \dots) \in F(H)$ yields zero,
    i.e. $a^-\{f\} \rhovac = 0$ for any $f \in H^*$.
	\item A symmetrized pure tensor $\SymmOperator_n (v_1 \otimes \dots \otimes v_n) \in  \SymmOperator_n\FunctionSpace^{\otimes n}$ with $v_i \in H$ is generated from a sequence of creation operators acting on $\rhovac$:
	\begin{equation}
    \SymmOperator_n (v_1 \otimes \dots \otimes v_n) = \aps{v_1} \dots \aps{v_n} \rhovac \,.
	\end{equation}

	\item The creation and annihilation operators satisfy the following commutation relations:
	\begin{subequations}
	\label{eq:a+a-CommutRels}
	\begin{align}
		[a^-\{f\},a^+\{w\}] & = \langle f,w\rangle \Id, \\
		[a^+\{w\},a^+\{\nu \}] &= 0, \\
		[a^-\{f\},a^-\{g\}] &= 0,
	\end{align}
	\end{subequations}
	for $w, \nu \in H$ and $f, g\in H^*$ and using the commutator $[a,b]:=ab-ba$ of operators $a, b$ on $\Fockspace$. The proofs are found in \cref{app:commutrels}.

	\item The particle number operator is defined as
	\begin{align}
		\mathcal{N}:= \sum_{\alpha} a^+\{u_\alpha\} a^-\{u_\alpha^*\},
	    \label{eq:partnumOperatorApp}
	\end{align} 
	where $(u_\alpha)_{\alpha\in \mathbb{N}}$ is a basis of the single-particle space $H$. The name of the operator refers to the fact that a density $\rho_n$ with fixed particle number $n$ is an \rev{eigenfunction} to the eigenvalue $n$, that is
	\begin{equation}
		\mathcal{N}\rho_n = n\rho_n , \quad \rho_n \in \SymmOperator_n H^{\otimes n}.
	\end{equation}
	We prove this statement in \cref{app:partnumoperator}.
For a general, normalized \rev{function} $\rho(t)  \in \Fockspace$ with $\| \rho(t) \|=1$, the average number of particles is obtained as:
\begin{equation}
  \| \mathcal{N} \rho(t) \| = \sum_{n\geq 0} \|\mathcal{N} \rho_n(t) \|_n
  = \sum_{n\geq 0} n P(n,t) = \Expect{N(t)} \,.
\end{equation}
Powers of $\mathcal{N}$ yield the higher-order factorial moments of $N(t)$, for example,
\begin{equation}
  \|\mathcal{N}^2\rho(t) \|= \Expect{N(t) (N(t) - 1)} \,,
\end{equation}
as shown in \cref{app:secondpartnumoperator}.
We note that \rrev{$\|\rho\|=1$ does not necessarily imply $\mathcal{N} \rho \in \Fockspace$ and the domain of $\mathcal{N}$ is only a subspace of $\Fockspace$. In particular,} there are reaction networks \rrev{showing an explosion in finite time $T<\infty$,} i.e.,
$\|\mathcal{N} \rho(t)\| = \mathbb{E}[N (t)] \to \infty$ for $t \to T$ although $\|\rho(t)\|=1$ for all $0 \leq t \leq T$.

\end{enumerate}

\section{Chemical diffusion master equation in terms of creation and annihilation operators} \label{sec:rep_operators}

\rev{The chemical diffusion-master equation (CDME) is generally composed of diffusion and reaction operators, see \cref{eq:emptyCDME}.
In this section, we will show how to formulate the CDME in a systematic way by expressing these operators in terms of creation and annihilation operators.}
We distinguish operators that conserve the number of particles and those that do not. The diffusion operator belongs to the first class, whereas the reaction operator of a general single reaction can be decomposed into a conserving and a non-conserving part.
\rev{We start with particle-number conserving operators and then follow up with reaction operators.}
The scheme is worked out in detail for two exemplary reaction systems at the end of this section. 


\subsection{Particle-number conserving operators}\label{sec:conserving}
We begin by expanding operators that do not change the number of particles if applied to a \rev{function} $\rho_n \in \SymmOperator_n H^{\otimes n}$, i.e., one with a determined number of particles. One can think of their action as moving particles in space or extracting information from the system. These operators will look different depending on the number of particles they act on. As an example, consider a many-particle system where particles diffuse independently. In this case, the diffusion operator is a conserving operator acting on single particles. Conversely, diffusion in the presence of pair interactions can be expressed by a conserving operator acting on two particles at a time.

\subsubsection{Conserving single-particle operators} \label{sec:conserving_single}

For operators acting on a single particle, let $A$ denote a linear operator on the single-particle space $H$. We denote by $\mathcal{A}_j$ the action of $A$ on the $j$th argument of a function in the many-particle space $H^{\otimes n}$. The action of $A$ on every particle of an $n$-particle system ($n \geq 1$) is then given by
\begin{align}
	\mathcal{A}^n := \sum_{j=1}^n \mathcal{A}_j,
	\label{eq:consvPartMulti}
\end{align}
and focusing on symmetrized spaces, one has $\mathcal{A}^n: \SymmOperator_n H^{\otimes n} \to \SymmOperator_n H^{\otimes n}$.
For any operator $A:H\to H$ 
it is shown in \cref{app:conervsingleoperator} that the operator $\mathcal{A}^n$ can be written as
\begin{align}
	\mathcal{A}= \sum_{\alpha ,\beta} \langle u_\alpha^*, A u_\beta\rangle \,\, \aps{u_\alpha} \ams{u_\beta^*},
	\label{eq:singPartExpansion}
\end{align}
where we dropped the superindex $n$ as the form of the right-hand side is the same for every $n$. Note that if $A$ is the identity, we recover the particle number operator $\mathcal{N}$, see \cref{eq:partnumOperatorApp}.

Applying the operator $\mathcal{A}$ component-wise extends its action to general elements of the Fock space, $\rho = (\rho_0,\rho_1,\rho_2,\dots)\in F(H)$, that is
$\mathcal{A}\rho := (\mathcal{A} \rho_0,\mathcal{A} \rho_1,\mathcal{A} \rho_2,\dots )$, with the convention that $\mathcal{A} \rho_0 := 0$.

\subsubsection{Conserving two-particle operators}

Analogously, let $B$ denote a linear operator on the two-particle space $H^{\otimes 2}$. The action of $B$ on every possible pair of particles of an $n$-particle system ($n \geq 2$) is
\begin{align}
	\mathcal{B}^n := \sum_{1\leq i< j \leq n} \mathcal{B}_{ij},
	\label{eq:cons2partOpMulti}
\end{align} 
where $\mathcal{B}_{ij}$ denotes the action of $B$ on the $i$th and $j$th components of a symmetrized many-particle function:
\begin{equation} \label{def:B_ij}
	\mathcal{B}_{ij}( \SymmOperator_n(v_1\otimes \dots \otimes v_n)) :=   \SymmOperator_n(B(v_i \otimes v_j)\otimes v_{\setminus \{i,j\}})
\end{equation}
where $v_{\setminus \{i,j\}}:=v_1\otimes \dots \otimes v_{i-1}\otimes v_{i+1}\otimes  \dots \otimes  \dots \otimes v_{j-1}\otimes v_{j+1}\otimes  \dots \otimes v_n$ for $i < j$.  We further assume $B$ to be symmetric, $\mathcal{B}_{i,j}=\mathcal{B}_{j,i}$, so it does not distinguish the labeling of the particles. We prove in \cref{app:conervtwooperator} that $\mathcal{B}^n$ has the following expansion:
\begin{align}
	\mathcal{B} 	&  =\frac{1}{2} \sum_{\substack{\alpha \leq \beta \\ \gamma \leq \delta}} \frac{1}{ s_{\gamma\delta}} \langle \SymmOperator_2(u_\alpha\otimes u_\beta)^*, B  (u_\gamma \otimes u_\delta)\rangle \,\, \aps{u_\alpha} \aps{u_\beta} \ams{u_\gamma^*} \ams{u_\delta^*} 
	\label{eq:twoPartExpansion3} \\
	&	 =\frac{1}{2}\sum_{\substack{\alpha , \beta \\ \gamma ,\delta}} \langle u_\alpha^* \otimes u_\beta^*, B (u_\gamma \otimes u_\delta)\rangle \,\, \aps{u_\alpha} \aps{u_\beta} \ams{u_\gamma^*} \ams{u_\delta^*}
	\label{eq:twoPartExpansion} 
\end{align}
for $s_{\alpha\beta}$ defined in \cref{eq:symm-coeff}.
Note that we again dropped the super index $n$ on $\mathcal{B}$ since the representations do not depend on $n$. The factor $1/2$ appears in the derivation, it accounts for the fact that removing first particle $i$ and then $j$ is the same as removing particle $j$ first and then~$i$.
Expansion \eqref{eq:twoPartExpansion3} is more adequate to easily obtain the explicit coefficients of complex expansions, e.g. three-particle operators ($n=3$). However, throughout this work, we will write the expansions in the form of \cref{eq:twoPartExpansion} due to its simpler notation. Also note that due to \cref{eq:symm-orth}, the expansion \eqref{eq:twoPartExpansion3} only needs the symmetrization operator in one of the arguments of the dual pairing.

We can again extend the action of the operator $\mathcal{B}$ to general functions in the Fock space, $\rho = (\rho_0,\rho_1,\rho_2,\dots)\in F(H)$, by applying the operator component-wise according to $\mathcal{B}\rho := (\mathcal{B} \rho_0,\mathcal{B} \rho_1,\mathcal{B} \rho_2,\dots ). $ Note $\mathcal{B} \rho_0 = \mathcal{B} \rho_1 = 0$ by construction.

Analogously, we could extend this result to conserving $n$-particle operators for $n \geq 3$; the coefficient leading the expansion would then be $1/n!$. However, in practice, it will be unlikely to encounter these operators for $n \geq 3$, so we do not explicitly work them out here.

\subsection{Reaction operators}
\label{sec:reaction_operators}

In the previous section, we focused on conserving operators that yield a result in $\SymmOperator_n H^{\otimes n}$ if acting on $\SymmOperator_n H^{\otimes n}$. However, reactions involve a change in the number of particles, and thus, they cannot be expressed only in terms of particle-number conserving operators. For an arbitrary reaction, the reaction operator can naturally be decomposed into two parts:
\begin{itemize}
	\item a conserving operator to \rev{specify} the probability outflow from the current state,
	\item a non-conserving operator to \rev{specify} the probability inflow from another state (with a different number of particles) into the current state.
\end{itemize}


In order to specify the two operators that form the reaction operator \rev{of a general reaction $ k\Species\rightarrow l \Species$}, we first need to define their action on a group of $k$ particles. Using the \rev{corresponding} reaction rate function \rev{$\lambda$ (see \cref{sec:intuitiveCDME})}, we introduce these operators as $\Lambda^{(k)}:H^{\otimes k} \to H^{\otimes k}$ and $\Lambda^{(k,l)}:H^{\otimes k} \to  H^{\otimes l}$ with
\rev{\begin{align}
	\left(\Lambda^{(k)}(u_{\beta_1}\otimes\dots\otimes u_{\beta_k})\right) (x^{(k)}) &:= (u_{\beta_1}\otimes\dots\otimes u_{\beta_k}) (x^{(k)})\int_{\SpaceOfMotion^l} \Propensity(y^{(l)},x^{(k)}) dy^{(l)}, \label{eq:propOperators_k}\\
	\left(\Lambda^{(k,l)}(u_{\beta_1}\otimes\dots\otimes u_{\beta_k})\right) (y^{(l)}) &:= \int_{\SpaceOfMotion^k} \Propensity(y^{(l)},x^{(k)})(u_{\beta_1}\otimes\dots\otimes u_{\beta_k}) (x^{(k)}) dx^{(k)}. \label{eq:propOperators_kl}
\end{align}}
Here, $\Lambda^{(k)}$ encodes the rate of leaving a given \rev{configuration $x^{(k)}$ with $k$ particles and going to any other configuration with $l$ particles,} 
whereas $\Lambda^{(k,l)}$ is the rate of going from \rrev{any} configuration of $k$ particles to \rrev{the} configuration \rev{$y^{(l)}$} with $l$ particles.

From the previous subsection, we already know how to extend a conserving operator to act on all possible combinations of $k$ particles in an $n$-particle space. For instance, for the reaction $\Species+\Species\rightarrow \Species$, the conserving part of the operator is [\cref{eq:twoPartExpansion}]:
\begin{equation}
	\ReactionOperator^{(2)}= \frac{1}{2}\sum_{\substack{\alpha_1, \alpha_2\\ \beta_1,\beta_2}} \langle u_{\alpha_1}^* \otimes u_{\alpha_2}^*, \Lambda^{(2)}( u_{\beta_1} \otimes u_{\beta_2} )\rangle \,\, \aps{u_{\alpha_1}} \aps{u_{\alpha_2}} \ams{u_{\beta_1}^*} \ams{u_{\beta_2}^*}.
\end{equation}
Analogously, we can construct a similar expression for the non-conserving operator, where two particles are removed and one particle is created:
\begin{equation}
	\ReactionOperator^{(2,1)} = \frac{1}{2}\sum_{\substack{\alpha\\ \beta_1,\beta_2}}  \langle u_\alpha^*, \Lambda^{(2, 1)} (u_{\beta_1} \otimes u_{\beta_2})  \rangle \aps{u_\alpha} \ams{u_{\beta_1}^*} \ams{u_{\beta_2}^*}.
\end{equation}
The proofs of the expansion of non-conserving operators are analogous to those for conserving operators given in  \cref{app:conervsingleoperator,app:conervtwooperator}. The reaction operator $\ReactionOperator$ in the Fock space is composed by the sum of these two operators, acting pointwise according to $(\ReactionOperator\rho)_n=  \ReactionOperator^{(2,1)}\rho_{n+1}-\ReactionOperator^{(2)}\rho_n$. 
Note that the conserving operator will always carry a minus in the CDME, as it refers to the outflow from a given state.
The joint action of both operators ensures that probability is conserved in the reaction.

For a general reaction $k\Species\rightarrow l\Species$, the reaction operator is given componentwise by $(\ReactionOperator\rho)_n=  \ReactionOperator^{(k,l)}\rho_{n+k-l}-\ReactionOperator^{(k)}\rho_n$, where
\begin{align}
	\ReactionOperator^{(k)}&= \frac{1}{k!}\sum_{\substack{\alpha_1, \dots , \alpha_k \\ \beta_1, \dots , \beta_k}} \left\langle \bigotimes_{i=1}^k u_{\alpha_i}^* , \Lambda^{(k)} \bigotimes_{j=1}^k u_{\beta_j} \right\rangle \,\,  \prod_{i=1}^{k} \aps{u_{\alpha_i}}\prod_{j=1}^{k} \ams{u_{\beta_j}^*}, \label{Rk}\\
	\ReactionOperator^{(k,l)} &= \frac{1}{k!}\sum_{\substack{ \alpha_1 ,\dots ,\alpha_l \\ \beta_1 , \dots , \beta_k}} \left\langle   \bigotimes_{i=1}^l u_{\alpha_i}^* , \Lambda^{(k,l)} \bigotimes_{j=1}^k u_{\beta_j} \right\rangle \prod_{i=1}^{l} \aps{u_{\alpha_i}}\prod_{j=1}^{k} \ams{u_{\beta_j}^*}. \label{Rkl}
\end{align}
Note that it holds $\ReactionOperator^{(k)}\rho_n = \ReactionOperator^{(k,l)}\rho_n = 0$ for $n<k$, \rev{which implies that $(\ReactionOperator\rho)_n= -\ReactionOperator^{(k)}\rho_n$ for $k \leq n< l$, while   $(\ReactionOperator\rho)_n= \ReactionOperator^{(k,l)}\rho_{n+k-l}$ for $l \leq n< k$}. As there are $k!$ ways to choose (i.e., remove) the same $k$ particles, the factor $1/k!$ is required to avoid double counting. Note these expansions can also be written in terms of the symmetrized basis, see \cref{app:reactionoperators}.

Below we list the expansions of the non-conserving part of frequently used reaction operators involving one species:
\begin{subequations}
	\begin{align}
		\ReactionOperator^{(0,1)} &= \sum_{\alpha} \langle u_\alpha^*, \Lambda^{(0, 1)} \mathbf{1} \rangle \aps{u_\alpha},\label{R01} \\
		\ReactionOperator^{(1, 0)} &= \sum_{\beta} \langle\mathbf{1} ,\Lambda^{(1, 0)} u_\beta \rangle \ams{u_\beta^*}, \label{R10} \\
		\ReactionOperator^{(0, 2)} &= \sum_{\alpha_1 , \alpha_2} \langle  u_{\alpha_1}^* \otimes u_{\alpha_2}^*, \Lambda^{(0,2)} \mathbf{1} \rangle \aps{u_{\alpha_1}} \aps{u_{\alpha_2}}, \\ 
		\ReactionOperator^{(2,0)} &= \frac{1}{2}\sum_{\beta_1 , \beta_2} \langle \mathbf{1}, \Lambda^{(2, 0)}(u_{\beta_1}\otimes u_{\beta_2}) \rangle \ams{u_{\beta_1}^*} \ams{u_{\beta_2}^*}, \\
		\ReactionOperator^{(1, 2)} &= \sum_{\alpha_1 ,\alpha_2} \sum_{\beta} \langle   u_{\alpha_1}^* \otimes u_{\alpha_2}^*, \Lambda^{(1, 2)} u_\beta  \rangle \aps{u_{\alpha_1}} \aps{u_{\alpha_2}} \ams{u_\beta^*}, \\
		\ReactionOperator^{(2,1)} &= \frac{1}{2}\sum_{\alpha} \sum_{\beta_1 ,\beta_2} \langle u_\alpha^*, \Lambda^{(2, 1)} (u_{\beta_1} \otimes u_{\beta_2})  \rangle \aps{u_\alpha} \ams{u_{\beta_1}^*} \ams{u_{\beta_2}^*}.
	\end{align}
\end{subequations}
These reaction operators are useful to formulate the CDME for a general reaction system.

\subsection{\rev{Representation of} the chemical diffusion master equation}
\label{sec:subCDME}

Given a set of reactions indexed by $r=1,..,S$, each of the form $k\mbox{A} \to l \mbox{A}$ for some $k,l \in \mathbb{N}_0$ and with corresponding reaction operators $\ReactionOperator_r$ composed by their conserving part $\ReactionOperator_r^{(k)}$ and their non-conserving part $\ReactionOperator_r^{(k,l)}$, the CDME \eqref{eq:emptyCDME} 
\rev{may be written as
\begin{equation}\label{eq:CDME}
\partial_t \rho = \left(\mathcal{D}  +  \sum_{r=1}^S \Big[\ReactionOperator_r^{(k,l)}-\ReactionOperator_r^{(k)}\Big]\right) \rho,
\end{equation}
where $k$ and $l$ depend on $r$.
}
For each component  $\rho_n$ of $\rho \in \Fockspace$ this means
\begin{equation}\label{eq:CDME_n}
	\partial_t \rho_n = \mathcal{D}\rho_n  +  \sum_{r=1}^S \left[\ReactionOperator_r^{(k,l)} \rho_{n+k-l} - \ReactionOperator_r^{(k)} \rho_{n}\right].
\end{equation}
That is, for several reactions, the overall reaction operator is simply the sum of individual reaction operators. 

\rev{
The representation \cref{eq:CDME} of the CDME, together with the reaction operator expansions in \cref{Rk,Rkl} form the main results of this work. It allows for a structured and systematic formulation of the probabilistic evolution described by the CDME with the physics of the system encoded in the given reaction rate functions and diffusion operators.
We note that the expansions of the conserving single and two-particle operators in \cref{eq:singPartExpansion,eq:twoPartExpansion}, respectively, are further useful to expand the diffusion operator or higher-order particle-conserving operators.

In \eqref{eq:CDME}, the operators do not explicitly depend on the particle number $n$, which greatly simplifies the expressions. With this, the operators $\AltReactionOperator_{nm}$ entering the matrix in \cref{eq:mainMasterEq} are given as
\begin{equation} \label{eq:Qmatrix}
	\AltReactionOperator_{nm} =
	\begin{cases}
		\sum_{r=1}^S \ReactionOperator_r^{(k,l)} \delta_{m-n,k-l} & \text{if} \quad m \neq n, \\
		-\sum_{r=1}^S \ReactionOperator_r^{(k)} & \text{if} \quad m = n.
	\end{cases}
\end{equation}
}
\rrev{%
The $\mathcal{Q}$-matrix has a band-structure since there is only a finite number of reactions, each of them changing the number of particles by a finite amount, i.e., there is some $\Delta n_\text{max}$ such that $\mathcal{Q}_{nm} = 0$ for $|n-m| > \Delta n_\text{max}$.
Another important property is related to the local conservation of probability and generalizes the fact that for a Markov chain on a finite state space column sums of the rate matrix are zero.
For the CDME, we show in \cref{sec:dissipative} that this is expressed for any fixed $m\in \mathbb{N}_0$ as
\begin{equation}
  \label{eq:conservation}
  \sum_{n=0}^\infty \mathcal{J}_{nm}(\cdot) = 0
\end{equation}
in terms of the linear functionals 
\begin{equation}
  \mathcal{J}_{nm}(\eta) := \int_{\SpaceOfMotion^m} (\AltReactionOperator_{nm} \eta)(x^{(m)}) \, dx^{(m)} \,,
  \quad \eta \in \SymmOperator_m H^m.
\end{equation}
As an interpretation, $\mathcal{J}_{nm}(\cdot)$ yields the probability flux from the space with $m$ particles to one with $n$ particles for $n \neq m$;
the total loss from the $m$-particle space is summarized in $-\mathcal{J}_{mm}(\cdot)$.
}

\rrev{
\paragraph*{CDME as an evolution equation.}
The CDME \eqref{eq:CDME}, shortly denoted as $\partial_t \rho = (\mathcal{D} + \ReactionOperator)\rho$ has the form of a linear evolution equation, and the question arises whether it is well-posed in the sense that solutions exist (in a finite time horizon, at least) and are unique. Furthermore, it is \emph{a priori} not clear that the positivity and normalization of the probability density are preserved, $\|\rho(t)\| = 1$ and $\rho_n(t) \geq 0$ for $t \geq 0$ and for each $n$, and how a permissible initial density $\rho(0)$ looks like.
These questions are brought up already when considering the scenario of well-mixed reaction-diffusion systems characterized by the conventional CME, which is an evolution equation on the space $\ell^1$ of discrete probability distributions $p=(p_0, p_1, \dots)$, where $p_n(t) = \|\rho_n(t)\|_n$.
The issue was addressed by the pioneers of probability theory for a broad class of Markov jump processes on a countable state space \cite{Feller:TAMS1940,Kolmogorov:1951,Doob:Stochastic}, providing non-trivial criteria to answer these questions \cite{Anderson:ContinuousTimeMarkovChains}.
Pathological examples can be found in refs.~\cite{Kendall:QJM1956,Reuter:AM1957,Anderson:ContinuousTimeMarkovChains} and the topic is still a subject of ongoing research \cite{Chen:AAP2004,Feinberg:AOR2017}.

Technically, one employs the Hille--Yosida theory in the setting of an abstract Banach space, which is $\Fockspace$ in our case, to specify necessary and sufficient conditions that the operator $\mathcal{A} := \mathcal{D}+\ReactionOperator$ generates a contraction semigroup
$\bigl(\exp(t \mathcal{A})\bigr)_{t\geq 0}$ on $F(H)$.
For a dissipative operator, the Lumer--Phillips theorem \cite{engel2000one,staffans2005well,schnaubelt2020lecture} provides more tractable criteria for this to hold. For example, in addition to verifying that $\mathcal{A}$ is dissipative on a suitable dense domain, which is proven in \cref{sec:dissipative}, one would need to investigate when the operator $\mu - \mathcal{A}$ is surjective for each $\mu > 0$.
A comprehensive analysis of the issue for the CDME exceeds the scope of the present work and is left for future research.
}


\paragraph*{Multiple species.}

Using the formalism developed above, we can obtain the CDME systematically for an arbitrary reaction-diffusion system with one chemical species.
Extensions to multiple species require a straightforward generalization of the notation: In the first place, for a system of $L$ species, the probability densities $\rho_n \in \SymmOperator_n H^{\otimes n}$ generalize to $\rho_{n_1,...,n_L} \in \SymmOperator_{n_1} H^{\otimes n_1} \otimes \dots \otimes \SymmOperator_{n_L} H^{\otimes n_L}$ with an index $n_l$ for the number of particles of each species; the arguments of $\rho_{n_1,...,n_L}$ refer to the particle positions for each species.
Note that since any two particles of different species are distinguishable, the $\rho_{n_1,...,n_L}$ must not be symmetrized with respect to all particle positions, but only with respect to those of each species.
The basis of the space of $\rho_{n_1,...,n_L}$ is the tensor product of the bases of the $n_1$- to $n_L$-particle spaces of the single species case.
Secondly, the diffusion part hardly changes except for the fact that the diffusion constant and physical interactions may vary among species, i.e., the diffusion operator will have as many indices as the system has species, indicating its different action on particles of different species.
Thirdly, for the reaction part, we get two indices per species, counting the gains or losses with respect to that species on the reactant as well as on the product side of the reaction. As a consequence, we will have rate functions $\Propensity(x^{(l_1)},\ldots x^{(l_L)};x^{(k_1)},\ldots x^{(k_L)})$ which depend on $2L$ position tuples for $L$ species. In the many-species generalization, the conserving part of the reaction operator will involve an integration over all product positions of all species. Analogously, the non-conserving part integrates over the reactant positions of all species. Also the creation and annihilation operators have to be specified as to which species it is that is created or destroyed, so that there are creation and annihilation operators separately for each species.

\subsection{Exemplary reaction systems} \label{sec:chemdiffusiveeq}

For the examples given in the following, we will reside with the setting of a single species for the sake of clarity.
\rev{Given the diffusion properties and the reaction rate functions of the system, which encode the physics, we will specify the CDME for some exemplary reaction systems in terms of creation and annihilation operators, and we will show that the equations are consistent with well-known results, e.g., those given in \cref{sec:intuitiveCDME}.} Any chemical reaction can be decomposed into a combination of unimolecular reactions and/or bimolecular reactions, we thus focus on creation and degradation to represent unimolecular reactions and on mutual annihilation as an example for a bimolecular reaction.

\subsubsection{Creation and degradation}
We consider the birth-death process given already in \cref{eq:degradation_creation_reactions}. It consists of the degradation and creation reactions, $\Species \to \emptyset$ and $\emptyset \to \Species$, occurring with rate functions $\Propensity_d(x)$ and $\Propensity_c(x)$, respectively;
in addition, each particle of species $\Species$ diffuses freely with diffusion coefficient $D$.
The chemical diffusion-master equation can be written in terms of a diffusion operator $\mathcal{D}$ and one reaction operator per reaction, $\ReactionOperator_d$ and $\ReactionOperator_c$, as shown in \cref{sec:reaction_operators}:
\begin{align}
	\partial_t \rho_n = \mathcal{D} \rho_n + \underbrace{\ReactionOperator_d^{(1,0)} \rho_{n+1}-\ReactionOperator_d^{(1)} \rho_n }_{\ReactionOperator_d \text{ (degradation)}} +\underbrace{ \ReactionOperator_c^{(0,1)} \rho_{n-1}-\ReactionOperator_c^{(0)}\rho_n }_{\ReactionOperator_c \text{ (creation)}},
	\label{eq:eom_creatdegrad}
\end{align}
with
\begin{subequations}
	\label[eqset]{eq:R1_birth_death}
	\begin{align}
		\mathcal{D} &= \sum_{\alpha ,\beta} \langle u_\alpha^*, D\nabla ^2 u_\beta\rangle \aps{u_\alpha} \ams{u_\beta^*},&&  \\
		\ReactionOperator_d^{(1)} &= \sum_{\alpha ,\beta} \langle u_\alpha^*, \Lambda^{(1)}_d u_\beta\rangle \aps{u_\alpha} \ams{u_\beta^*},
		&\ReactionOperator_d^{(1,0)} &= \sum_{\alpha} \langle \mathbf{1} ,\Lambda^{(1, 0)}_d  u_\alpha \rangle \ams{u_\alpha^*},  \\
		\ReactionOperator_c^{(0)} &= \Lambda^{(0)}_c,
		&\ReactionOperator_c^{(0,1)} &= \sum_{\alpha} \langle u_\alpha^*, \Lambda^{(0, 1)}_c \mathbf{1} \rangle \aps{u_\alpha},
	\end{align}
\end{subequations}
where $\rho_n = \rho_n(x^{(n)},t)$, and the reaction rate operators for each reaction are
\begin{subequations}
	\begin{align}
		\left(\Lambda^{(1)}_du_{\beta}\right) (x) &=  \Propensity_d(x)u_{\beta} (x),
		&\Lambda^{(1,0)}_d u_{\alpha} &= \int_\SpaceOfMotion \Propensity_d(x)u_{\alpha} (x) dx, \label{Lambda_deg}  \\
		\Lambda^{(0)}_c &= \int \Propensity_c(y)dy,
		& \left(\Lambda^{(0,1)}_c \mathbf{1} \right) (x) &= \Propensity_c(x),\label{Lambda_cre}
	\end{align}
\end{subequations}
see \cref{eq:propOperators_k,eq:propOperators_kl}.
Writing the CDME in matrix form as in \cref{eq:mainMasterEq}, the reaction operator
\begin{equation}
	\ReactionOperator =
	\begin{pmatrix}
		-\ReactionOperator_c^{(0)} & \ReactionOperator_d^{(1,0)} & 0 & 0 & \dots & \\
		\ReactionOperator_c^{(0,1)} & -(\ReactionOperator_d^{(1)} + \ReactionOperator_c^{(0)}) & \ReactionOperator_d^{(1,0)} & 0 &  \dots &  \\
		0 & \ReactionOperator_c^{(0,1)} & -(\ReactionOperator_d^{(1)} + \ReactionOperator_c^{(0)}) & \ReactionOperator_d^{(1,0)} & \dots \\
		\vdots & \vdots & & \ddots
	\end{pmatrix}
\end{equation}
attains a tridiagonal form. Along the diagonal there is the conserving part $\ReactionOperator_d^{(1)} + \ReactionOperator_c^{(0)}$ with a minus sign, while the non-conserving parts $\ReactionOperator_d^{(1,0)}$ and $\ReactionOperator_c^{(0,1)}$ are found on the secondary diagonals. Moreover, as the reaction operators given in terms of creation and annihilation operators do not depend on the particle number $n$, we let them act directly on elements of the Fock space, $\rho=(\rho_0,\rho_1,\dots,\rho_n,\dots)$, and re-write the CDME in compact form:
\begin{equation}
	\partial_t \rho = \left(\mathcal{D}  + \ReactionOperator_d^{(1,0)} - \ReactionOperator_d^{(1)} + \ReactionOperator_c^{(0,1)} -
	\ReactionOperator_c^{(0)}\right) \rho \,.
\end{equation}

To write the CDME explicitly as in \cref{sec:CDME}, we start with the particle-number conserving operators. The expansions in terms of creation/annihilation operators corresponds to applying the operators to all the possible particles the operators can act on, thus
\begin{align}
	\mathcal{D}\rho_n &= \sum_{i=1}^n D\nabla_i^2  \rho_n, \qquad
	\ReactionOperator_d^{(1)}\rho_n = \sum_{i=1}^n \Propensity_d(x_i) \rho_n, \qquad
	\ReactionOperator_c^{(0)}\rho_n = \rho_n \int_{\SpaceOfMotion} \Propensity_c(y)dy, \label{DR1R0_rho}
\end{align}
where the last one corresponds to a conserving zero-particle operator and thus it acts on no particles. For the non-conserving operators, we need to explicitly apply the creation and annihilation operators. In \cref{app:apam_densities_proof}, we show how to apply the creation and annihilation operators from \cref{def:longsymaj+-} to densities. Using these relations and using
$\rho_n=\sum_{\beta_1\leq \dots \leq \beta_n}\tilde c_{\beta_1,\dots,\beta_n} \SymmOperator_n (u_{\beta_1} \otimes \dots \otimes u_{\beta_n})$, we obtain
\begin{align}
	(\ReactionOperator_d^{(1,0)} \rho_{n+1})(x^{(n)}) & \stackrel{\eqref{R10} }{=}  \sum_{\beta} \langle \mathbf{1} ,\Lambda^{(1, 0)}_d  u_\beta \rangle \left( a^-\{u^*_\beta\} \rho_{n+1}\right)(x^{(n)}) \nonumber\\
	& \stackrel{\eqref{def:densityaj-2}}{=}  \sum_{\beta} \langle \mathbf{1} ,\Lambda^{(1, 0)}_d  u_\beta \rangle (n+1)\int_\SpaceOfMotion u^*_\beta(y) \rho_{n+1}(x^{(n)},y) dy  \nonumber \\
	&  \stackrel{\eqref{eq:resunity}}{=} (n+1) \int_\SpaceOfMotion \Propensity_d(y) \rho_{n+1}(x^{(n)},y) dy \label{R10_rho}
\end{align}
and
\begin{align}
	(	\ReactionOperator_c^{(0,1)}\rho_{n-1})(x^{(n)}) & \stackrel{\eqref{R01} }{=}  \sum_{\alpha} \langle u_\alpha^*, \Lambda_c^{(0, 1)} \mathbf{1} \rangle\left( a^+\{u_\alpha\}\rho_{n-1}\right)(x^{(n)}) \nonumber\\
	&\stackrel{\eqref{def:densityaj+2}}{=} \sum_{\alpha} \langle u_\alpha^*, \Lambda_c^{(0, 1)} \mathbf{1} \rangle \frac{1}{n} \sum_{j=1}^n u_\alpha (x_j^{(n)}) \rho_{n-1}(x^{(n)}_{/j}) \nonumber\\
	&\stackrel{\eqref{eq:expansionHelement}}{=} \frac{1}{n} \sum_{j=1}^n  \Propensity_c(x_j^{(n)}) \rho_{n-1}(x^{(n)}_{/j}). \label{R01_rho}
\end{align}
Gathering all the terms, \rev{this matches exactly the proposal in \cref{eq:time_evolution}, which serves as a first consistency check of the presented formalism.}

\subsubsection{Mutual annihilation}
Next, we consider the mutual annihilation reaction $\Species + \Species \to\emptyset$ with rate function $\Propensity (x_1,x_2)$.  The CDME in component-wise form is simply
\begin{equation} \label{eq:CDME_mutual}
	\partial_t \rho_n = \mathcal{D}\rho_n  + \ReactionOperator^{(2,0)}\rho_{n+2} - \ReactionOperator^{(2)}\rho_n
\end{equation}
for $n\geq 0$.
The diffusion term is the same as before, and the reaction operator can be decomposed into two parts, as shown in \cref{sec:reaction_operators}:
\begin{subequations}
	\label[eqset]{eq:R2_mutual}
	\begin{align}
		\ReactionOperator^{(2)} &= \frac{1}{2}\sum_{\substack{\alpha_1,\alpha_2 \\ \beta_1,\beta_2}} \langle u_{\alpha_1}^* \otimes u_{\alpha_2}^*, \Lambda^{(2)} (u_{\beta_1} \otimes u_{ \beta_2}) \rangle  \aps{u_{\alpha_1}} \aps{u_{\alpha_2}} \ams{u_{\beta_1}^*} \ams{u_{\beta_2}^*} \,, \label{R2}  \\
		\ReactionOperator^{(2,0)} &= \frac{1}{2}\sum_{\beta_1 , \beta_2} \langle \mathbf{1}, \Lambda^{(2, 0)}(u_{\beta_1}\otimes u_{\beta_2}) \rangle \ams{u_{\beta_1}^*} \ams{u_{\beta_2}^*} \,.
	\end{align}
\end{subequations}
In matrix notation, the full reaction operator acting on $\rho \in \Fockspace$ reads
\begin{equation}
	\ReactionOperator =
	\begin{pmatrix}
		-\ReactionOperator^{(2)} & 0 & \ReactionOperator^{(2,0)} & 0 & 0 & \dots & \\
		0 & -\ReactionOperator^{(2)} & 0 & \ReactionOperator^{(2,0)}& 0 &  \dots &  \\
		\vdots & \vdots & &  \ddots &
	\end{pmatrix} \,.
\end{equation}

We will now recover the explicit form of the CDME for mutual annihilation. For the conserving part of the reaction operator,
we use the relations \cref{eq:cons2partOpMulti,def:B_ij,eq:twoPartExpansion} and apply definition \eqref{eq:propOperators_k}
of the reaction rate operator to obtain
\begin{align}
	\ReactionOperator^{(2)}\rho_n = \sum_{1\leq i < j \leq n} \Propensity(x_i^{(n)}, x_j^{(n)}) \rho_n. \label{R2_rho}
\end{align}
For the non-conserving operator, we insert the definitions of the reaction rate operators [\cref{eq:propOperators_kl}] and of the annihilation operator [\cref{def:longsymaj-}],
\begin{align}
	\ReactionOperator^{(2,0)} \rho_{n+2}
	&= \frac{1}{2}\sum_{\beta_1,\beta_2} \langle \mathbf{1}, \Lambda^{(2, 0)}(u_{\beta_1}\otimes u_{\beta_2}) \rangle \rrev{\sum_{\substack{i,j=1\\j\neq i}}^{n+2} \langle u_{\beta_1}, v_i \rangle \langle u_{\beta_2}, v_j \rangle \SymmOperator_{n} (v_{\setminus \{i,j\}})}\nonumber \\
	&=	\frac{1}{2}\sum_{\beta_1,\beta_2} \langle \mathbf{1}, \Lambda^{(2, 0)}(u_{\beta_1}\otimes u_{\beta_2}) \rangle \rrev{\sum_{\substack{i,j=1\\j\neq i}}^{n+2} \langle u_{\beta_1} \otimes u_{\beta_2}, v_i \otimes v_j  \rangle
	\SymmOperator_{n} (v_{\setminus \{i,j\}})} \nonumber \\
	&=	\frac{1}{2}\rrev{\sum_{\substack{i,j=1\\j\neq i}}^{n+2} \langle  \mathbf{1},  \Lambda^{(2, 0)} (v_i \otimes v_j)  \rangle \SymmOperator_{n} (v_{\setminus \{i,j\}})}
 \nonumber\\
	&=	\frac{1}{2}\rrev{\sum_{\substack{i,j=1\\j\neq i}}^{n+2}}   \int_{\SpaceOfMotion^2} \Propensity(y_1,y_2) \rho_{n+2}(x^{(n)}, y_1, y_2)dy_1 dy_2 \nonumber\\
	&=	\frac{(n+2) (n+1)}{2} \int_{\SpaceOfMotion^2} \Propensity(y_1, y_2) \rho_{n+2}(x^{(n)},y_1,y_2)dy_1 dy_2 \label{R20_rho}
\end{align}
\rrev{for $v_{\setminus \{i,j\}}:=v_1\otimes \dots \otimes v_{i-1}\otimes v_{i+1}\otimes  \dots \otimes  \dots \otimes v_{j-1}\otimes v_{j+1}\otimes  \dots \otimes v_{n+2}$}.
So, the full expression of the CDME of mutual annihilation is given in component-wise form as
\begin{multline} \label{eq:CDME_mutual_int}
	\partial_t \rho_n(x^{(n)}) =  \mathcal{D}\rho_n(x^{(n)})  + \frac{n(n-1)}{2}  \int_{\SpaceOfMotion^2} \Propensity(y_1,y_2) \rho_{n+2}(x^{(n)},y_1,y_2)dy_1 dy_2 \\
	-\sum_{1\leq i < j \leq n} \Propensity(x_i^{(n)}, x_j^{(n)}) \rho_n(x^{(n)}) \,.
\end{multline}
\rev{By comparing to the CME for mutual annihilation \cite{winkelmann2020stochastic}, one finds that both the structure of the equation and the coefficients obtained within the present approach are consistent with well-known models at a more coarse-grained level.} 
Note that for many analyses and derivations, this explicit form of the equation is not needed. Instead, we will work on the operator level in terms of creation and annihilation operators in the following.

\section{Spatial discretization} \label{Spatial_dis}

\rev{In the previous sections we laid out a theoretical basis for the probabilistic description of particle-based reaction-diffusion systems. Adapting the concept of the Fock space, we developed a systematic method to formulate CDMEs for arbitrary reaction networks. The CDME fully characterizes the stochastic reaction-diffusion dynamics. Solving it analytically or numerically, however, will in general be a demanding issue.} It either requires solving directly a huge system of partial differential equations (PDEs) or
performing stochastic simulations of the underlying particle-based reaction-diffusion process to obtain costly Monte Carlo estimates of the solution to the CDME.
\rev{However, there are situations where the physical or structural features of a specific system permit an approximation of the solutions by a (small) finite set of distinguished basis functions.
In such cases, a significant complexity reduction is achieved by a projection onto the subspace spanned by this reduced basis set.
Our framework is an ideal starting point for this as it already provides a basis representation of the CDME. As an example, we will use a basis of indicator functions completely covering the domain of particle positions, which amounts to a spatial coarse-graining. We thus obtain a generalized RDME ---consistent with the convergent RDME \cite{isaacson2013convergent,isaacson2018unstructured}--- that extends the standard RDME by reactions between particles located in different subdomains.
}

\subsection{Galerkin projection} \label{sec:copynumber}

Let $\hat H \subset H$ be a finite-dimensional linear subspace of $H$, and let $\xi_1,\dots,\xi_M$ be a normed basis of $\hat H$, i.e., $\|\xi_i\|_1=1$ for $i=1,\dots,M$. 
The dual basis $\xi_1^*,\dots,\xi_M^*$ is such that $\langle \xi_i^*,\xi_j \rangle = \delta_{i,j}$
for all $i, j$.
For example, the $\xi_i$ could be indicator functions of subsets for a given spatial discretization---a special case which  will be analyzed in \cref{STCME}.
The set $\hat H= \Span( \xi_i )$ induces a subspace $\hat{F}\subset F(H)$ of the Fock space $F(H)$ defined in \cref{def:fockspace}.
We will now consider a projection $Q:F(H) \to \hat{F} \subset F(H)$ onto this subspace.

The Galerkin ansatz for an element $\rho\in F(H)$ is given by $\hat \rho =(\hat \rho_0,\hat \rho_1,\dots)\in \hat{F}$ with
\begin{equation}\label{hat_rho_n}
	\hat \rho_n = Q\rho_n= \sum_{1 \leq i_1\leq \ldots \leq i_n\leq M}  c_{i_1,\ldots,i_n} \,
	\SymmOperator_n(\xi_{i_1}\otimes\ldots\otimes\xi_{i_n})
\end{equation}
for coefficients
\begin{align}
	c_{i_1,\ldots,i_n}  &:= \langle \SymmOperator_n  (\xi_{i_1} \otimes \dots \otimes \xi_{i_n})^*,\rho_n\rangle  \nonumber \\
	&~= \frac{1}{s_{\xi_{i_1},\dots,\xi_{i_n}}}\langle \SymmOperator_n  (\xi^*_{i_1} \otimes \dots \otimes \xi^*_{i_n}),\rho_n\rangle
\end{align}
with $s_{\xi_{i_1},\dots,\xi_{i_n}}$ defined in \cref{eq:symm-coeff}. In accordance with the notation in \cref{sum_notation}, we write
\begin{align}\label{basis_repr}
	\hat\rho  & = \sum_{n=0}^\infty \sum_{i_1\leq \ldots \leq i_n}  c_{i_1,\ldots,i_n} \,
	\SymmOperator_n(\xi_{i_1}\otimes\ldots\otimes\xi_{i_n})  \in \hat{F}
\end{align}
for the basis representation, or, in terms of creation operators,
\begin{align}\label{basis_repr2}
	\hat\rho  & = \sum_{n=0}^\infty \sum_{i_1\leq \ldots \leq i_n}  c_{i_1,\ldots,i_n} \,
	a^+\{\xi_{i_1}\} \dots a^+\{\xi_{i_n}\} \,\rhovac,
\end{align}
where
$\rhovac = (1, 0, 0, \dots) \in F(H)$ is the normalised vacuum \rev{element}.
Equal indices are again included in the sum because, e.g., $\xi_1 \otimes \xi_1$ is an allowed two-particle \rev{density}.
In the following, we will use the abbreviations
\begin{equation}
	a_i^+ := a^+ \{\xi_i \} \quad \text{and} \quad
	a_i^- := a^- \{\xi_i^* \} \,.
\end{equation}

For the diffusion operators $\mathcal{D}$ and the reaction operators $\ReactionOperator$ we derive the projected operators $\hat{\mathcal{D}}=Q\mathcal{D}Q$ and $\hat{\ReactionOperator}=Q\ReactionOperator Q$, respectively, by extending the basis of the subspace $\hat{H}$ by  the complement basis $\chi_1,\chi_2,\dots$, such that
$(u_1,u_2,\dots) = (\xi_1,\dots, \xi_M, \chi_1,\chi_2,\dots)$ is a basis of the full space $H$.
Then, using the operator expansions derived in \cref{sec:rep_operators}, we obtain equivalent expressions for the projected operators: After projecting, only the sum over the $\xi_i$ components remains due biorthogonality, i.e.,
$\langle \chi_j^*, \xi_i \rangle = \langle \xi_i^*, \chi_j \rangle = 0$ for any $i, j$.
The expressions for the projected operators can thus be obtained by simply replacing sums over $(u_\alpha)$ by sums over $(\xi_i)$.
For example, for the diffusion operator $\mathcal{D}$, we obtain
\begin{equation}\label{D_projected}
	\hat{\mathcal{D}}
	= \sum_{i,j} \langle \xi_i^* ,\mathcal{D}\xi_j \rangle a^+_i a^-_j ,
\end{equation}
see \cref{eq:singPartExpansion}, while for the non-conserving operator $\ReactionOperator^{(0,1)} $ of the reaction $\emptyset \to A$, we get
\begin{equation}
	\hat{\ReactionOperator}^{(0,1)} = \sum_{i}\langle \xi_i^*, \Lambda^{(0, 1)} \mathbf{1} \rangle a^+_i
\end{equation}
instead of the original expression \cref{R01}.

\paragraph*{Copy number representation.} 

The copy number representation introduced in \cref{eq:copynumBasisExp} is particularly useful for \rev{densities} projected onto a finite Galerkin basis $\xi_1,\ldots, \xi_M$.
We simply need to redefine the index set $\mathbb{M}_n$ of \cref{M_n} by restricting to the $M$ basis elements, i.e., to multi-indices $N=(N_1,\dots,N_M)$. Analogously to \cref{eq:copynumBasisExp}, the $n$-particle \rev{density} $\hat{\rho}_n$ as in \cref{hat_rho_n} can then be written as
\begin{align} \label{eq:rho_galerkin}
	\hat\rho_n = & \sum_{N\in\mathbb{M}_n }  p_{N_1,\ldots,N_M} \,
	\SymmOperator_n(\xi_1^{\otimes N_1}\otimes \dots \otimes\xi_M^{\otimes N_M}) \nonumber \\
	= & \sum_{N\in\mathbb{M}_n }  p_{N_1,\ldots,N_M} \,
	(a_1^+)^{N_1}\ldots(a_M^+)^{N_M}\rhovac
\end{align}
with coefficients
\begin{align} \label{p_N}
	p_{N_1,\ldots, N_M} &= \frac{n!}{N_1! \dots N_M!} \left\langle \xi_{1}^{^*\otimes N_1} \otimes \dots \otimes \xi_{M}^{^*\otimes N_M}, \rho_n\right \rangle.
\end{align}
In the following sections we will use the short-hand notation
\begin{equation} \label{copynumber}
	\ket{N_1, \dots, N_M} := (a_1^+)^{N_1}\ldots(a_M^+)^{N_M}\rhovac
\end{equation}
for the copy number representation.
An element of the projected Fock space $\hat{F}$ then has the copy number representation
\begin{align}
	\hat\rho = & \sum_{n\geq0}\sum_{N\in \mathbb{M}_n}
	p_{N_1,\ldots,N_M} \, \ket{N_1, \dots, N_M} \nonumber \\
	= & \sum_{N_1,\dots,N_M=0}^\infty
	p_{N_1,\ldots, N_M}  \, \ket{N_1, \dots, N_M}, \label{Eq:Galerkin-B}
\end{align}
replacing the representation \eqref{basis_repr2}.

\subsection{Normalization and positivity of the projected densities}

Given $\rho\in F(H)$ with $\rho_n\geq 0$ for all $n$ and $\|\rho\|=1$, we would like to get a projected function $\hat \rho$ which fulfills these properties, too.
In general, using \cref{norm} and \cref{n-norm} as well as the standard triangle inequality, we have
\begin{equation}
	\|\hat \rho \| \leq  \sum_{n=0}^\infty \sum_{N\in\mathbb{M}_n }  |p_{N_1,\ldots,N_M}| \cdot
	\|\SymmOperator_n(\xi_1^{\otimes N_1}\otimes \dots \otimes\xi_M^{\otimes N_M})\|.
\end{equation}
From $\|\xi_i\|_1=1$ for all $i$, it follows that $\|\xi_1^{\otimes N_1}\otimes \dots \otimes\xi_M^{\otimes N_M}\|_n=1$ as well as $\|\SymmOperator_{n}(\xi_1^{\otimes N_1}\otimes \dots \otimes\xi_M^{\otimes N_M})\|_n=1$ for all $N=(N_1,\dots,N_M)\in \mathbb{M}_n$, such that we obtain
\begin{equation}
	\|\hat \rho \| \leq  \sum_{n=0}^\infty  \sum_{N\in\mathbb{M}_n }  |p_{N_1,\ldots,N_M}|.
\end{equation}
In order to obtain equality, we have to assume both the basis functions $\xi_i$ and the coefficients $p_{N_1,\ldots,N_M}$ to be positive, in which case it holds $\hat\rho_n\geq 0$ for all $n$ and
\begin{equation}\label{equality}
	\|\hat \rho \| = \sum_{n=0}^\infty \sum_{N\in\mathbb{M}_n } p_{N_1,\ldots,N_M}.
\end{equation}
This will be true for the special case of a full spatial partition with rescaled indicator functions, as we will see below.

However, as soon as the basis functions $\xi_i$ are positive-valued and overlapping in \rev{position} space $\SpaceOfMotion$, the dual basis functions $\xi_i^*$ have negative values and, consequently, also the coefficients $p_{N_1,\ldots,N_M}$ as defined by the dual pairing \eqref{p_N} can become negative. In this case, equality \eqref{equality} is not fulfilled and an interpretation of the coefficients as probabilities becomes pointless.

The positivity of the coefficients is only guaranteed if we assume the dual basis functions $\xi_i^*$ to be positive, $\xi_i^*\geq 0$ for all $i$.  Demanding in addition that the dual basis functions sum up to one everywhere in \rev{position} space, $\sum_i \xi_i^*(x^{(1)}) =1$ for all $ x^{(1)} \in \SpaceOfMotion$ (as given, e.g., for committor functions in the context of transition path theory \cite{weinan2006towards,metzner2009transition}), this guarantees that the coefficients sum up to one, as well. In this case, the functions $\xi_i$ themselves are not necessarily positive, and consequently, also the projected functions $\hat{\rho}_n$ can have negative values, which renders their interpretation as projected probability distribution questionable.  These insights motivate us to restrict the following analysis to a full-partition approach.

\subsection{Full-partition approach}

In order to make sure that both the coefficients and the projected probability distributions are positive-valued, we consider non-overlapping basis functions as indicator functions of a full spatial partition into subsets.
That is, we split the \rev{position} space $\mathbb{X}$ into finitely many disjoint subsets $\mathbb{X}_i$, $i=1,\dots,M$ such that
\begin{equation}
 \mathbb{X} = \bigcup_{i=1}^M \mathbb{X}_i, \quad \mbox{with } \mathbb{X}_i\cap \mathbb{X}_j = \emptyset \mbox{ for } i\neq j.
\end{equation}
Given these subsets $\mathbb{X}_i\subset \mathbb{X}$,  we can define the rescaled indicator functions
\begin{equation}\label{rescaled_indicator}
	\xi_i:=\frac{1}{\mbox{vol}(\mathbb{X}_i)}1_{\mathbb{X}_i}=\frac{1}{\|1_{\mathbb{X}_i}\|_1}1_{\mathbb{X}_i} \,,
\end{equation}
which fulfill  $\|\xi_i\|_1=1$. In this special case, the duals are given by
\[ \xi^*_i= \mbox{vol}(\mathbb{X}_i)\cdot \xi_i=1_{\mathbb{X}_i}.\]

For this approach of a full spatial partition, the norm is naturally retained: Let $\rho \in F(H)$ be such that $\rho_n \geq 0$ for all $n$ and $\|\rho\|=1$ for the norm defined in \cref{norm}. This means that it holds
\begin{equation}\label{sum_P_N}
	\sum_{n=0}^\infty\mathbb{P}[N=n] =1
\end{equation}
for $\mathbb{P}[N=n] := \int \rho_n(x^{(n)}) \, dx^{(n)} \geq 0,$
see also \cref{P(N)}. Now we note that for fixed $n$  it holds
\begin{equation}\label{sum_p_N}
	\sum_{N\in \mathbb{M}_n} p_N = \sum_{i_1,\dots,i_n} \langle \xi_{i_1}^* \otimes \ldots \otimes \xi_{i_n}^*, \rho_n \rangle
\end{equation}
for the coefficients $p_N=p_{N_1,....,N_M}$ defined in \cref{p_N}.
Moreover, we have
\begin{equation}
	\sum_{i_1,\dots,i_n} (\xi_{i_1}^* \otimes \ldots \otimes \xi_{i_n}^*)(x^{(n)}) = \sum_{i_1,\dots,i_n} \xi^*_{i_1}(x^{(n)}_1) \cdot \ldots \cdot \xi^*_{i_n}(x^{(n)}_2) = 1 \quad \forall x^{(n)} \in \mathbb{X}^n
\end{equation}
for the given basis functions $\xi_i$.
Inserting into  \cref{sum_p_N} we obtain
\begin{align}
	\sum_{N\in \mathbb{M}_n} p_N &= \sum_{i_1,\dots,i_n} \int (\xi_{i_1} \otimes \ldots \otimes \xi_{i_n})^*(x^{(n)}) \rho_n(x^{(n)}) \, dx^{(n)} \nonumber \\
	&=   \int \rho_n(x^{(n)}) \, dx^{(n)} \nonumber \\
	&= \mathbb{P}[N=n],
\end{align}
so, due to property \eqref{sum_P_N}, we find
\begin{equation}
	\|\hat \rho \| = \sum_{n=0}^\infty \sum_{N\in \mathbb{M}_n} p_N =1
\end{equation}
with $p_N \geq 0$ for all $N$. This means that the norm is preserved and the coefficients $p_N$ may be interpreted as probabilities. Also $\hat{\rho}_n\geq 0$ is fulfilled such that $\hat{\rho}_n$ is again a probability density function. \rrev{This discretization can be understood in analogy to conservative finite volume methods, where the conserved quantity is the probability.}

\paragraph*{Average concentration field.}
As a simple application of the Galerkin projection and the probabilistic interpretation of \cref{p_N}, we calculate the expected number of particles in a ball $B_\epsilon(y)$ of radius $\epsilon$ centred at $y \in \mathbb{X}$. For the Galerkin basis, we choose only $M=2$ basis functions, namely the indicator functions on $\mathbb{X}_1 = B_\epsilon(y)$ and on its complement $\mathbb{X}_2 = \mathbb{X} \setminus B_\epsilon(y)$.
Then, the probability that at time $t$ there are $n$ particles in the system and exactly $k$ out of them are located in $B_\epsilon(y)$ is obtained from the $n$-particle component of a given $\rho(t) \in \Fockspace$ as the coefficient $p_{k,n-k}(t)$ in \cref{p_N}, explicitly:
\begin{equation}
	p_{k,n-k}(t) = \frac{n!}{k!(n-k)!} \int_{B_\epsilon(y)^k \times (\mathbb{X}\setminus B_\epsilon(y))^{n-k}} \, \rho_n(x^{(n)},t) \, dx^{(n)} \,.
\end{equation}
The average number of particles in $B_\epsilon(y)$ is found by taking the expectation of $k$ with respect to the joint distribution of $k$ and $n$ as
\begin{equation}
	X_\epsilon(y, t) := \sum_{n=0}^\infty \sum_{k=1}^n k \, p_{k,n-k}(t) .
\end{equation}
Dividing further by the volume of the ball and taking its radius to zero yields the average molecular concentration at point $y$:
\begin{equation}
	c(y, t) := \lim_{\epsilon \to 0} \frac{1}{\mathrm{vol}(B_\epsilon(y))} X_\epsilon(y, t) \,,
\end{equation}
which, after carrying out the limit, reduces to
\begin{align}
	c(y, t) &= \sum_{n=0}^\infty \sum_{k=1}^{n} \frac{n!}{(k-1)!(n-k)!} \int_{\{y\}^{k-1} \times \mathbb{X}^{n-k}} \, \rho_n(y, x^{(n-1)},t) \, dx^{(n-1)} \nonumber \\
	&= \sum_{n=0}^\infty n \int_{\mathbb{X}^{n-1}} \, \rho_n(y, x^{(n-1)},t) \, dx^{(n-1)}
	\label{eq:conc}
\end{align}
since in the first line the integral is non-zero only for $k=1$ as $\{y\}^{k-1}$ is a null set otherwise.
The last result, \cref{eq:conc}, can serve as starting point to derive the deterministic reaction-diffusion equation in terms of average concentrations from a given CDME.

\subsection{Generalized reaction-diffusion master equation}\label{STCME}

Given the setting of a full spatial partition with basis functions defined in  \cref{rescaled_indicator} we will investigate next the projected diffusion and reaction operators for some basic scenarios and derive the corresponding generalized RDME as an evolution equation for the coefficients $p_N$. Explicit calculations can be found in \cref{app:diff-reac-ops}.

\subsubsection{Diffusion}
Consider the projected diffusion operator given in \cref{D_projected} and define $d_{ij} := \left\langle \xi_i^*, \mathcal{D}\xi_j \right\rangle$, such that
\begin{equation} \label{D_projected2}
	\hat{\mathcal{D}} = \sum_{i,j} d_{ij}a^+_i a^-_j .
\end{equation}
This means that $d_{ij}$ refers to the rate to go from basis $j$ to basis $i$. 
We apply this operator $\hat{\mathcal{D}}$ to $\hat \rho$ given in \cref{Eq:Galerkin-B}. We find
(see \cref{eq:Dhat})
\begin{multline}
	\hat{\mathcal{D}} \hat{\rho}
	= \sum_{N_1,\dots,N_M} \sum_{\substack{i,j\\i\neq j}}
	p_{N_1,\ldots ,N_i-1, \ldots, N_j+1, \ldots , N_M} d_{ij}
	(N_j+1)\cdot
	\ket{N_1,\ldots,N_M} \\
	+ \sum_{N_1,\dots,N_M} \sum_{i}
	p_{N_1,\ldots ,N_M}d_{ii}  N_i\cdot \ket{N_1,\ldots,N_M}.
\end{multline}
Defining the operator $\mathrm{D}$ acting on the coefficients according to
\begin{align}
	\mathrm{D}p_{N_1,\ldots ,N_M} & := \sum_{\substack{i,j\\i\neq j}}p_{N_1,\ldots, N_i-1, \ldots, N_j+1, \ldots , N_M} d_{ij}
	(N_j+1) +  \sum_{i}	p_{N_1,\ldots ,N_M}d_{ii}  N_i 	\nonumber \\
	& = \sum_{\substack{i,j\\i\neq j}}\left[ d_{ij}
	(N_j+1)p_{N_1,\ldots ,N_i-1, \ldots, N_j+1, \ldots , N_M} - d_{ji} N_i  	p_{N_1,\ldots ,N_M} \right] , \label{Dc}
\end{align}
we can write
\begin{equation}
	\hat{\mathcal{D}} \hat{\rho} =
	\sum_{N_1,\dots,N_M} Dp_{N_1,\ldots ,N_M} \cdot
	\ket{N_1, \dots, N_M}.
\end{equation}
In \cref{Dc}  we used the identity $d_{ii}= - \sum_{j\neq i} d_{ji}$, which follows for the chosen basis of indicator functions from the equality $\xi^*_i =1-\sum_{j\neq i} \xi^*_j$ after integration by parts.

In the time-dependent setting, where $\rho=\rho(\cdot,t)$ is a function of $t$, we turn the diffusion equation $\partial_t \rho = \mathcal{D} \rho$ into a
``diffusion master equation'' for the time-dependent coefficients $p_{N_1,\dots,N_M}(t)$ of $\hat  \rho (t)$ and get
\begin{align}
	\frac{d}{dt}p_{N_1,\dots,N_M}(t)  & = 	\mathrm{D}p_{N_1,\ldots, N_M}(t)
\end{align}
as the diffusive part of the RDME.

\subsubsection{Creation and degradation}

Consider the chemical diffusion-master equation for creation and degradation, see \cref{eq:eom_creatdegrad}. Let $\Propensity_d$ denote the rate function of the degradation reaction  $\Species \to \emptyset$, while $\Propensity_c$ is the rate function for the creation reaction $\emptyset \to \Species$. The corresponding operators are denoted by $\ReactionOperator^{(1)}_d$ and  $\ReactionOperator^{(1,0)}_d$ for degradation and by $\ReactionOperator^{(0)}_c$ and  $\ReactionOperator^{(0,1)}_c$ for creation.

\paragraph*{Degradation.}

The projected conserving one-particle reaction operator $\hat{\ReactionOperator}_d^{(1)}$ acts similarly to the diffusion operator. Defining
\begin{equation}\label{def:lambda}
	\Propensity_d^{i} := \left\langle \xi^*_i, \Lambda_d^{(1)}\xi_i\right\rangle = \int \xi^*_i(x)\Propensity_d(x)\xi_i(x) dx = \int \Propensity_d(x)\xi_i(x) dx
\end{equation}
for $\Lambda_d^{(1)}$ given in \cref{Lambda_deg},
we find [\cref{eq:Rd1_hat}]:
\begin{equation}
	\hat{\ReactionOperator}_d^{(1)} \hat\rho = \sum_{N_1,\dots,N_M} \sum_{i}
	p_{N_1,\ldots, N_M}\Propensity_d^{i}  N_i\cdot
	\ket{N_1,\ldots,N_M}.
\end{equation}
Applying the projected non-conserving operator $\hat{\ReactionOperator}_d^{(1,0)}= \sum_i \langle 1,\Lambda_1^{(1,0)}\xi_i \rangle a^-_i$ to $\hat \rho = \sum_{n \geq 0}\hat \rho_n$ given in \cref{Eq:Galerkin-B}, we get [\cref{eq:Rd10_hat}]
\begin{equation}
	\hat{\ReactionOperator}_d^{(1,0)} \hat{\rho}
	=  \sum_{N_1,\dots,N_M} \sum_i
	p_{N_1,\ldots, N_i+1, \ldots ,N_M} \,\Propensity_d^i  (N_i+1)\cdot
	\ket{N_1,\ldots,N_M}.
\end{equation}
Combining the results, we define the operator $\mathrm{R}_{d}$ of the degradation reaction in terms of coefficients  by

\begin{equation}\label{operator_degradation}
	\mathrm{R}_d p_{N_1,\dots,N_M} := 	\sum_i \left[\Propensity_d^i  (N_i+1) p_{N_1,\ldots, N_i+1 \ldots, N_M}\,- \Propensity_d^i  N_i p_{N_1,\ldots ,N_M} \right].
\end{equation}

\paragraph*{Creation.}

As for the creation reaction $\emptyset \to \Species$, the conserving part $\hat{\ReactionOperator}_c^{(0)}$  acts as a mere constant
[\cref{eq:Rc0_hat}]:
\begin{align}
	\hat{\ReactionOperator}_c^{(0)} \hat\rho &= \Lambda_c^{(0)} \sum_{N_1,\dots,N_M}
	p_{N_1,\ldots, N_M} \ket{N_1,\ldots,N_M} \nonumber \\
	&= \Lambda_c^{(0)}  \hat\rho \,,
\end{align}
with the reaction rate constant
\begin{equation}\label{def:lambda_0}
	\Lambda_c^{(0)} = \int \Propensity_c(x) \, dx.
\end{equation}
For the nonconserving part $\hat{\ReactionOperator}^{(0,1)}$ we set
\begin{equation}\label{def:lambda_01}
	\Propensity_c^i := \langle \xi^*_i,\Lambda_c^{(0,1)}1 \rangle = \int \xi^*_i(x) \,\Propensity_c(x) \,dx
\end{equation}
and obtain [\cref{eq:Rc01_hat}]
\begin{align}
	\hat{\ReactionOperator}_c^{(0,1)} \hat\rho
	&= \sum_{N_1,\dots,N_M} \sum_i
	p_{N_1,\ldots, N_i -1,\ldots , N_M}\Propensity_c^i   \,
	\ket{N_1,\ldots,N_M} .
\end{align}
In terms of coefficients, and observing that $\Lambda_c^{(0)} =\sum_i \Propensity_c^i$, we obtain the operator $\mathrm{R}_{c}$ as
\begin{align} \label{eq:R_creation}
	\mathrm{R}_{c} p_{N_1,\dots,N_M} := &
	\sum_i  \Propensity_c^i \left[p_{N_1,\ldots,N_i-1,\ldots,N_M}-  p_{N_1,\ldots,N_M} \right].
\end{align}

In combination with the diffusion operator, we obtain the standard RDME for the system of creation and degradation
\begin{align} \label{eq:gRDME_birth_death}
	\frac{d}{dt}p_{N_1,\dots,N_M}(t)  &= (\mathrm{D} + \mathrm{R}_{d}+\mathrm{R}_{c})p_{N_1,\dots,N_M}(t) \nonumber \\
	&= \sum_{\substack{i,j\\i\neq j}}\left[ d_{ij}
	(N_j+1)p_{N_1,\ldots ,N_i-1, \ldots, N_j+1, \ldots , N_M}(t) - d_{ji} N_i  	p_{N_1,\ldots ,N_M}(t) \right] \nonumber \\
	& \qquad + 	\sum_i \Propensity_d^i\left[  (N_i+1) p_{N_1,\ldots, N_i+1, \ldots, N_M}(t) \,-   N_i p_{N_1,\ldots ,N_M}(t) \right] \nonumber \\
	& \qquad + \sum_i\Propensity_c^i \left[  p_{N_1,\ldots, N_i-1, \ldots, N_M}(t)-  p_{N_1, \ldots,  N_M}(t) \right].
\end{align}

\paragraph*{Scaling by volume.}
Assuming that the reaction rate function $\Propensity_c$ of creation is constant, $\Propensity_c(x)=\gamma$ for all $x\in \SpaceOfMotion$, it holds $\Propensity_c^i=\gamma \int \xi^*_i(x) dx= \gamma \mathrm{vol}(\mathbb{X}_i)$ and $\Propensity_c= \gamma \mathrm{vol}(\mathbb{X})$. As for degradation, we analogously obtain $\Propensity_d^i=\gamma$. In both cases, this is the usual volume-scaling of a reaction of order zero or one, respectively.

\subsubsection{Mutual annihilation}

For the  mutual annihilation given by the reaction $\Species + \Species \rightarrow \emptyset $ let $\Propensity(x^{(2)})$ denote the rate for two particles to react when they are located at $x^{(2)}=(x_1^{(2)},x_2^{(2)})\in \SpaceOfMotion^2$. By virtue of \cref{eq:propOperators_kl} with $k=2$ and $l=0$  we set
\begin{align}
	\Propensity^{ij} & := \langle \xi^*_i\otimes\xi^*_j,\Lambda^{(2)} (\xi_i\otimes\xi_j)\rangle \nonumber \\
	& =\int (\xi^*_i\otimes\xi^*_j)(x^{(2)})\Propensity(x^{(2)}) (\xi_i\otimes\xi_j)(x^{(2)})\, dx^{(2)} \label{lambda_mutual}
\end{align}
and obtain the conserving part $\hat{\ReactionOperator}^{(2)}$ of the reaction as [\cref{eq:R2_hat}]
\begin{multline}
	\hat{\ReactionOperator}^{(2)} \hat\rho
	=
	\sum_{N_1,\dots,N_M}
	\Bigg[
	\sum_{i}
	p_{N_1,\ldots, N_M}\frac{1}{2} \Propensity^{ii} N_i(N_i-1)\,
	\ket{N_1,\ldots,N_M} \\
	+ \sum_{i< j}
	p_{N_1,\ldots, N_M} \Propensity^{ij} N_i N_j \,
	\ket{N_1,\ldots,N_M}
	\Bigg].
	\label{eq:annih-C}
\end{multline}
On the other hand, under the assumption of non-overlapping basis functions, we find the same $\Propensity^{ij}$ for the nonconserving part:
\begin{equation}
	\langle (1,\Lambda^{(2,0)} (\xi_i\otimes\xi_j)\rangle=\int \Propensity(x^{(2)}) (\xi_i\otimes\xi_j)(x^{(2)})\, dx^{(2)} = \Propensity^{ij},
\end{equation}
and thus [\cref{eq:R20_hat}]
\begin{multline}\label{eq:annih-NC}
	\hat{\ReactionOperator}^{(2,0)} \hat\rho =
	\sum_{N_1,\dots,N_M} \Bigg[
	\sum_{i}
	p_{N_1,\ldots, N_i+2,\ldots, N_M} \frac{1}{2}\Propensity^{ii} (N_i+2)(N_i+1) \,
	\ket{N_1,\ldots,N_M} \\
	+\sum_{i<j}
	p_{N_1,\ldots, N_i+1,\ldots, N_j+1, \ldots, N_M} \Propensity^{ij} (N_i+1) (N_j+1) \, 	\ket{N_1,\ldots,N_M}\Bigg].
\end{multline}
In terms of the coefficients, we obtain the \textit{generalized RDME}
\begin{align} \label{eq:gRDME_mutual}
	\frac{d}{dt}&p_{N_1,\dots,N_M}(t) = (\mathrm{D}+\mathrm{R})p_{N_1,\dots,N_M}(t)  \nonumber \\
	&=	\sum_{\substack{i,j\\i\neq j}}\left[ d_{ij}
	(N_j+1)p_{N_1,\ldots ,N_i-1, \ldots, N_j+1, \ldots , N_M}(t) - d_{ji} N_i  	p_{N_1,\ldots ,N_M}(t) \right] \nonumber \\
	& \qquad +
	\frac{1}{2}\sum_i \Propensity^{ii} \left[(N_i+2)(N_i+1)p_{N_1,\ldots, N_i+2, \ldots, N_M}(t)   - N_i(N_i-1)  p_{N_1, \ldots,  N_M}(t) \right]
	\nonumber\\
	& \qquad +\sum_{i< j} \Propensity^{ij} \left[(N_i+1)(N_j+1)p_{N_1,\ldots, N_i+1, \ldots, N_j+1, \ldots, N_M}(t) - N_i N_jp_{N_1, \ldots,  N_M}(t) \right],
\end{align}
where the last two lines define the operator $\mathrm{R}$. The very last line refers to reactions between particles located in different "boxes". This makes the equation more general than the standard RDME which would only allow reactions between particles located in the same box.

\paragraph*{Volume scaling of the rate constant.}
Let the reaction rate function $\Propensity$ be of the form $\Propensity(x,y)=\gamma 1_{B_r(x)}(y)$,
which corresponds to the standard situation of the Doi reaction model \cite{doi1976stochastic};
here, $\gamma > 0$ is a reaction rate constant, $r > 0$ is some maximum reaction distance,
and $1_{B_r(x)}$ denotes the indicator function of the ball $B_r(x)$ of radius $r$ centered at $x$, i.e.,
$1_{B_r(x)}(y) = 1$ if $\|x - y\| \leq r$ and 0 otherwise.
For simplicity, we ignore boundary effects for positions $x$ close to a subdomain boundary such that the reactive volume around $x$ can be assumed to be the same all over a subdomain $\SpaceOfMotion_i$, i.e.,
$\int_{\SpaceOfMotion_i} 1_{B_r(x)}(y) dy = V_0$ does not depend on $x \in \SpaceOfMotion_i$,
which is justified if the volume of $\SpaceOfMotion_i$ is large compared to the reactive volume \cite{kostre2020coupling}.
Then, we obtain
\begin{align}
	\Propensity^{ii} & =  \int_{\SpaceOfMotion \times \SpaceOfMotion} \xi_i^*(x)\xi_i^*(y)\Propensity(x,y) \xi_i(x)\xi_i(y) dx dy \nonumber \\
	& = \frac{\gamma}{\mbox{vol}(\SpaceOfMotion_i)^2} \int_{\SpaceOfMotion_i \times \SpaceOfMotion_i} 1_{B_r(x)}(y) \,dx dy \nonumber \\
	& = \frac{\gamma}{\mbox{vol}(\SpaceOfMotion_i)^2} \int_{\SpaceOfMotion_i} V_0 dx  \nonumber \\
	& = \frac{\gamma V_0}{\mbox{vol}(\SpaceOfMotion_i)} ,
\end{align}
which is the standard scaling of the reaction rate constant for a second-order reaction in a finite volume.

\section{Discussion and perspectives}


In summary, we developed a probabilistic framework to \rev{formalize} stochastic, particle-resolved reaction-diffusion dynamics. 
A main result is the \rev{structured formulation of the} CDME \eqref{eq:CDME_n}, which governs the temporal evolution of the probability distribution of the many-particle open system, \rev{by means of systematically constructed diffusion and reaction operators}.
\rev{As the framework is based on suitable creation and annihilation operators,}
we translated the concept of Fock space, well-known for Hilbert spaces in quantum mechanics,
to the setting of probability densities, which are absolutely integrable rather than square-integrable.
We introduced creation and annihilation operators [\cref{def:longsymaj+-}] as a natural means to express operations on symmetrized many-particle densities, and we used them as basic building blocks \rev{to represent the reaction and diffusion operators, and thus the complete CDME. Diffusion operators are expressed as conserving particle operators, as they do not change the number of particles. However, reaction operators are decomposed into two parts, one that conserves and one that does not conserve the number of particles}, representing loss and gain contributions to the probability density, respectively [e.g., \cref{eq:R1_birth_death,eq:R2_mutual}].
At the core of the reaction operators are position-dependent reaction rate functions, which encode details of the underlying PBRD model such as where to place reaction products \cite{agmon1990theory, andrews2004stochastic, delrazo2016discrete, froehner2018reversible}. \rev{Domains with open boundary conditions can be simulated by including reactions that insert or remove particles in a boundary layer only.} Some of the reaction operators take explicitly the form of integrals over $n$-particle densities e.g., [\cref{R20_rho}],
so the CDME in general is a family of coupled integro-partial-differential equations.
\rrev{The CDME is proposed as an evolution equation on the Fock space $\Fockspace$. 
Similarly as for the CME -- valid for well-mixed systems -- there is the question when do solutions to the CDME exist and are unique, which will be addressed elsewhere.}
\rev{We worked out the explicit form of the CDME for typical examples, namely a birth-death process [\cref{eq:eom_creatdegrad}] and a mutual annihilation reaction [\cref{eq:CDME_mutual}]. We point out that the framework extends beyond these specific examples and is applicable to general reaction schemes.}
We briefly sketched the extension to multiple species (see \cref{sec:subCDME}), which is straightforward for few species, but results in a clumsy notation for a larger number of species.
Using the Galerkin projection technique to discretize space, we derived a generalized RDME with non-local higher-order reactions as an approximation of the CDME [e.g., \cref{eq:gRDME_birth_death,eq:gRDME_mutual}].
As a by-product, we found relations between the reaction rate functions of the CDME and the rate constants of the RDME [\cref{lambda_mutual}].


The (generalized) RDME as a spatial discretization of the CDME may, in principle, appear suitable for a direct numerical treatment
due to its form as an evolution equation for the expansion coefficients of the \rev{Fock space element} $\hat\rho \in \hat F$.
However, such an attempt would be computationally prohibitive, which follows from a brief estimate of the amount of data to be processed.
For a Galerkin basis of length $M$ and limiting the maximum copy number per basis element to $N_\mathrm{max}$,
corresponding to a maximum of $M \times N_\mathrm{max}$ molecules in the system,
the \rev{function} $\hat\rho$ has coefficients $p_{N_1,\dots,N_M}$ for $M$ indices $N_1, \dots, N_M$ each ranging from $1$ to $N_\mathrm{max}$, see \cref{eq:rho_galerkin}.
The memory requirement to store such a dense coefficient array is $(N_\mathrm{max})^M$ floating-point numbers,
which in the setting of the ST-CME with few compartments and an exemplary, conservative choice of $M=5$ and $N_\mathrm{max} = 100$ would require already $(N_\mathrm{max})^M=10^{10}$ double words or 75 gigabytes of memory that need to be processed in every time integration step.
Conversely, in the RDME-like setting of a regular mesh of $M=10\times 10$ cells with $N_\mathrm{max}=5$, one would need the unthinkable amount of $(N_\mathrm{max})^M\approx 10^{70}$ double words.
Thus instead of solving the CDME or RDME directly,
the method of choice is to calculate Monte Carlo estimates from simulations of the underlying stochastic process either
in continuous space for the CDME \cite{hoffmann2019readdy,dibak2019diffusion}
or, much less costly, on a lattice for the RDME \cite{drawert2012urdme,roberts2013lattice,hallock2014simulation}.

The obtained framework is broadly applicable for formulating and analyzing a wide range of arbitrary reaction-diffusion processes.
It provides the theoretical backbone to unify reaction-diffusion models at multiple scales that emerge from the CDME in specific limits or regimes.
\rev{For instance, macroscopic and mesoscopic descriptions rely on (possibly fluctuating) concentration fields.
Testing the assumptions behind the emergence of these descriptions requires analysis on a more refined model such as the CDME.} The present formalism also enables the development of consistent hybrid and multi-scale numerical schemes.
For example, we have shown that the spatial discretization of the CDME by a Galerkin projection
naturally yields a consistent, non-local RDME, which supports higher-order reactions between different subdomains
and which converges to the original CDME as the mesh size tends to zero---in contrast to the standard local RDME.
Our work thus complements previous numerical studies on convergent RDMEs \cite{isaacson2013convergent,isaacson2018unstructured}.

\rev{A relevant direction of future research would consist on fully integrating out the spatial degrees of freedom from the CDME, which should recover the classical CME under well-mixed conditions.}
This is straighforward for unimolecular reactions [\cref{eq:example_cme}],
whereas higher-order reactions are influenced by diffusion such that the spatial distribution of reacting molecules enters the rates effectively observed in the CME \cite{rice1985diffusion,hanggi1990reaction,agmon1990theory,dibak2019diffusion};
further it gives rise to additional reaction channels \cite{gopich2013diffusion, gopich2019diffusion} and non-Markovian effects \cite{gopich2018theory,grebenkov2018strong,froemberg2021generalized}.
Moreover, deterministic reaction-diffusion equations emerge from spatially resolved stochastic models in the limit of large copy numbers \cite{kostre2020coupling, isaacson2020mean,isaacson2020reaction}. It would be of practical value to derive such, in general non-linear, macroscopic equations from the CDME; see \cref{eq:conc} for how to obtain locally averaged molecular concentrations from the probability densities of the CDME.
Future research may also include the extension of \cref{Spatial_dis} to other types of spatial discretizations by projecting onto basis functions different from indicator functions, e.g., committor functions or locally supported, piecewise-linear functions (hat functions), yielding alternative RDME models that could be numerically advantageous.
In conclusion, we have given the probabilistic description of particle-resolved reaction-diffusion systems along with systematic means to formulate the corresponding CDME, which opens rich perspectives for future mathematical and numerical investigations.

\section*{Acknowledgements}
We thank Rupert Klein for a fruitful exchange on the manuscript.
We acknowledge the support of Deutsche Forschungsgemeinschaft (DFG) through the Collaborative Research Center SFB~1114 “Scaling Cascades in Complex Systems”, project no.~235221301, sub-project C03, and under Germany's Excellence Strategy -- MATH+ : The Berlin Mathematics Research Center (EXC-2046/1) -- project no.\ 390685689 (subproject AA1-1). MJR acknowledges support by the Dutch Institute for Emergent Phenomena (DIEP) cluster at the University of Amsterdam and support from DFG grant no. RA 3601/1-1 during the final stages of the project.

%


\appendix
\renewcommand{\thesubsection}{\thesection.\arabic{subsection}}

\section{Basic relations and properties}

Here, we prove some basic relations and properties for creation and annihilation operators that are quoted in the main text or are used for proving some of the subsequent results.

\subsection{Relations for the dual pairing}
\label{app:dualpairingsymmetric}

For the contraction of dual pairings [\cref{dual_pairing_1}], it holds for $\eta \in \FunctionSpace$ and $\zeta \in \FunctionSpace^*$ and an operator $A : \FunctionSpace \to \FunctionSpace$ that
\begin{align} \label{eq:resunity}
	\langle \zeta , A \eta\rangle
	&= \sum_{\alpha} \langle \zeta , A u_\alpha \rangle \langle u_\alpha^*, \eta \rangle \nonumber \\
	&= \sum_{\alpha} \langle \zeta , u_\alpha \rangle \langle u_\alpha^*, A \eta \rangle \,,
\end{align}
where we inserted the basis expansion of $\eta$ given in \cref{eq:expansionHelement}.
In the second line, the corresponding expansion of $A\eta \in \FunctionSpace$ was used.

Next, we prove \cref{eq:symm-orth}. Assuming $\zeta \in (H^{\otimes n})^*$ and $\eta \in H^{\otimes n}$, it holds
\begin{align}
	\langle \SymmOperator_n \zeta, \eta \rangle = \int_{\SpaceOfMotion^{n}} \SymmOperator_{n}(\zeta(x^{(n)})) \eta(x^{(n)})) d x^{(n)}.
\end{align}
As the integral is over the whole domain, the labeling of the integration variables does not matter, and consequently
\begin{align}
	\langle \SymmOperator_n \zeta, \eta \rangle &=
	\frac{1}{n!} \sum_{\sigma \in \Sigma_n}
	\int_{\SpaceOfMotion^{n}} \SymmOperator_{n}(\zeta(x^{(n)})) \eta( x^{(n)}_{\sigma (1)}, \dots, x^{(n)}_{\sigma (n)}) d x^{(n)} \notag\\
	&=\int_{\SpaceOfMotion^{n}} \SymmOperator_{n}(\zeta(x^{(n)})) \SymmOperator_{n}(\eta(x^{(n)})) d x^{(n)} =
	\langle \SymmOperator_n \zeta, \SymmOperator_n \eta \rangle,
\end{align}
where $\Sigma_n$ is the set of all permutations $\sigma$ on $\{1, \dots, n\}$.
Following the same argument in reverse gives
$
\langle \SymmOperator_n \zeta, \SymmOperator_n \eta \rangle
= \langle \zeta, \SymmOperator_n \eta \rangle
$
and thus
\begin{equation}
	\langle \SymmOperator_n \zeta, \eta \rangle
	= \langle  \zeta, \SymmOperator_n\eta \rangle \,.
\end{equation}

\subsection{Products of $a^+$ and $a^-$ operators}

In the following, we prove relations for the action of products of $a^+$ and $a^-$ operators on a symmetrized pure tensor $v=\SymmOperator_n (v_1 \otimes \cdots \otimes v_n)\in \SymmOperator_n H^{\otimes n}$ for $v_i \in H$ and $i=1, \dots, n$.

Applying $a^+ a^-$ for elements of the single-particle basis to $v$ yields
\begin{align}
	\aps{u_\alpha} \ams{u_\beta^*} v  =\SymmOperator_n   \sum_{j=1}^n \langle u_\beta^*, v_j\rangle ( u_\alpha \otimes v_{\setminus \{j\}}),
	\label{eq:apam_gen}
\end{align}
where we used the definitions \eqref{def:longsymaj+-}, as well as the short-hand notation $v_{\setminus \{j\}} = v_1\otimes \dots v_{j-1}\otimes v_{j+1} \dots \otimes v_n$.
Similarly, we find for the product $a^+ a^+ a^- a^-$ that
\begin{align}
	\aps{u_\alpha} \aps{u_\beta} &\ams{u_\gamma^*} \ams{u_\delta^*} v =\SymmOperator_{n}\sum_{\substack{i,j=1 \\ j \neq i}}^n \langle u_\delta^*,v_i \rangle   \langle u_\gamma^*,v_j \rangle (u_\alpha \otimes u_\beta \otimes v_{\setminus \{i,j\}}).
	\label{eq:apapamam_gen}
\end{align}

\subsection{Commutation relations}
\label{app:commutrels}
We prove the commutator relations \eqref{eq:a+a-CommutRels}.
Again, it is sufficient to consider a symmetrized pure tensor $v \in \SymmOperator_n H^{\otimes n}$ as above.
Let $w, \nu \in H$ and $f, g\in H^*$, and we begin with the first identity:
\begin{align}
	[a^-\{f\},&a^+\{w\}] v = a^-\{f\}a^+\{w\}v - a^+\{w\}a^-\{f\}v.
\end{align}
To expand the first term, we apply \cref{def:longsymaj+} followed by \cref{def:longsymaj-}. The second term is given in \cref{eq:apam_gen}, and we obtain
\begin{align}
	[a^-\{f\}, a^+\{w\}] v
	&= \SymmOperator_n \bigl(\langle f,w\rangle v + \sum_{j=1}^{n} \langle f,v_j\rangle (w \otimes v_{\setminus \{j\}}) \bigr)
	- \SymmOperator_n \sum_{j=1}^{n} \langle f,v_j\rangle ( w \otimes v_{\setminus \{j\}}) \notag \\
	&= \langle f, w\rangle v.
\end{align}
The other two commutators follow similarly by inserting the definitions \eqref{def:longsymaj+-}:
\begin{align}
	[a^+\{w\},a^+\{\nu \}]v &= a^+\{w\}a^+\{\nu \}v - a^+\{\nu \}a^+\{w\}v  \notag \\
	&=\SymmOperator_{n+2}(w\otimes \nu \otimes v_1 \otimes \dots\otimes v_n) - \SymmOperator_{n+2}(\nu \otimes w \otimes v_1 \otimes \dots\otimes v_n)  \notag \\
	&=0,  \\
	\intertext{and}
	[a^-\{f\},a^-\{g\}]v &= a^-\{f\}a^-\{g\}v - a^-\{g\}a^-\{f\}v  \notag \\
	&= \sum_{\substack{i,j \\ i\neq j}}\left( \langle f,v_i\rangle \langle g,v_j\rangle - \langle f,v_j\rangle \langle g,v_i\rangle \right)  \SymmOperator_{n-2}(v_{\setminus\{i,j\}})     \notag \\
	&=0 .
\end{align}

\subsection{Particle number operator}
\label{app:partnumoperator}
Putting $\alpha = \beta$ in \cref{eq:apam_gen}, summation over $\alpha$ yields the particle number operator:
\begin{align}
	\mathcal{N} v &= \sum_{\alpha} \aps{u_\alpha} \ams{u_\alpha^*} v  \notag \\
	&= \sum_{\alpha} \SymmOperator_n \sum_{j=1}^n   \langle u_\alpha^*, v_j\rangle u_\alpha \otimes v_{\setminus \{j\}}  \notag \\
	&= \sum_{j=1}^n \SymmOperator_n (v_1 \otimes \cdots \otimes v_n) \notag \\
	&= n v,
\end{align}
exploiting biorthogonality, $\langle u_\alpha^*, v_j\rangle = \delta_{\alpha,j}$, in the last but one step.
As a general density $\rho_n \in \SymmOperator_n H^{\otimes n}$ is a linear combination of symmetric pure tensors, the previous relation implies $\mathcal{N}\rho_n= n \rho_n$.
It follows straightforwardly that the identity operator (resolution of unity) on $\SymmOperator_n H^{\otimes n}$ reads
\begin{align}
	\mathcal{I} \rho_n  = \frac{1}{n} \sum_\alpha \aps{u_\alpha} \ams{u_\alpha^*} \rho_n  \,.
	\label{eq:resUnity1app}
\end{align}

\subsection{Second-order particle number operator}
\label{app:secondpartnumoperator}
Taking $\alpha = \gamma$ and $\beta = \delta$ in \cref{eq:apapamam_gen} and summing over $\alpha$ and $\beta$, we obtain the second-order particle number operator:
\begin{align}
	\mathcal{N}^2 v &= \sum_{\alpha\beta}
	\aps{u_\alpha} \aps{u_\beta} \ams{u_\alpha^*} \ams{u_\beta^*} v \nonumber \\
	&=\sum_{\alpha\beta} \SymmOperator_{n}\sum_{\substack{i,j=1 \\ j \neq i}}^n \langle u_\beta^*,v_i \rangle   \langle u_\alpha^*,v_j \rangle (u_\alpha \otimes u_\beta \otimes v_{\setminus \{i,j\}}) \nonumber \\
	&=\sum_{\substack{i,j=1 \\ j \neq i}}^n \SymmOperator_n (v_1 \otimes \cdots \otimes v_n) \nonumber \\
	&= n (n-1) v. \label{eq:partnumOperator2App}
\end{align}
Due to linearity, this implies $\mathcal{N}^2\rho_n= n(n-1) \rho_n$.
In analogy to the particle number operator it holds
\begin{align}
	\| \mathcal{N}^2 \rho \| = \sum_{n\geq 0} \|\mathcal{N}^2 \rho_n \|_n
	= \sum_{n\geq 0} n(n-1) P(n) = \Expect{N(t)(N(t)-1)}.
\end{align}
The generalization is straightforward. In general, $\mathcal{N}^k$ can be expressed as $k$ sums where each term is composed by $k$ creation operators followed by $k$ annihilation operators. It holds that
\begin{equation}
	\|\mathcal{N}^k \rho\| = \Expect{N(t)(N(t)-1)\cdots(N(t)-k+1)},
\end{equation}
which are the factorial moments of $N(t)$.

\section{Expansions in terms of creation and annihilation operators}
\label{app:proofsapamoperators}

In the following, we always assume $v \in \SymmOperator_n H^{\otimes n}$ to be a symmetrized pure tensor, $v=\SymmOperator_n (v_1 \otimes \cdots \otimes v_n)$ for $v_i \in H$.

\subsection{Creation and annihilation operators acting on densities }
\label{app:apam_densities_proof}
We rewrite the creation and annihilation operators from \cref{def:longsymaj+-} acting explicitly on symmetrized densities.
For $\rho_n \in \SymmOperator_n H^{\otimes n}$ and $w \in H$ it holds
\begin{align}
	a^+\{w\}\rho_{n} &\stackrel{\eqref{def:longsymaj+}}{=}  \SymmOperator_{n+1} (w \otimes \rho_n) \label{eq:aps_expansion_app_0} \\
	&\stackrel{\eqref{eq:expansionHelement}}{=} \sum_{\alpha} \langle u_\alpha^*, w \rangle \SymmOperator_{n+ 1}(u_\alpha \otimes \rho_n) \nonumber \\
	&= \sum_{\alpha} \langle u_\alpha^*, w \rangle a^+\{ u_\alpha \}\rho_{n} \,,\label{eq:aps_expansion_app}
\end{align}
which evaluates at a point $x^{(n+1)} \in \mathbb{X}^{n+1}$ to
\begin{align}
	(a^+\{w\}\rho_{n})(x^{(n+1)})
	= \frac{1}{n+1}\sum_{j=1}^{n+1} w(x_j^{(n+1)})\rho_n(x_{\setminus \{j\}}^{(n+1)}) \label{def:densityaj+2}
\end{align}
with the notation $x_{\setminus \{j\}}^{(n+1)} = (x_1^{(n+1)}, \dots, x_{j-1}^{(n+1)}, x_{j+1}^{(n+1)}, \dots x_{n+1}^{(n+1)}) \in \mathbb{X}^n$.

Applying the annihilation operator to $\rho_n=\sum_{\beta_1\leq \dots \leq \beta_n}\tilde c_{\beta_1,\dots,\beta_n} \SymmOperator_n (u_{\beta_1} \otimes \dots \otimes u_{\beta_n})$ yields for $f \in \FunctionSpace^*$:
\begin{align} \label{eq:ams_expansion_app}
	a^-\{f\}\rho_{n}
	&\stackrel{\eqref{def:longsymaj-}}{=} \sum_{j=1}^n
	\sum_{\beta_1\leq \dots \leq \beta_n} \langle f, u_{\beta_j} \rangle \tilde c_{\beta_1,\dots,\beta_n} \SymmOperator_{n-1} (u_{\setminus \{\beta_j\} }) \nonumber \\
	&\stackrel{\eqref{eq:expansionHelement}}{=} \sum_{j=1}^n
	\sum_{\beta_1\leq \dots \leq \beta_n} \sum_{\alpha} \langle f, u_\alpha \rangle \langle u_\alpha^*, u_{\beta_j} \rangle \tilde c_{\beta_1,\dots,\beta_n} \SymmOperator_{n-1} (u_{\setminus \{\beta_j\} }) \nonumber \\
	&\stackrel{\eqref{def:longsymaj-}}{=} \sum_{\alpha} \langle f, u_\alpha \rangle a^-\{u_\alpha^*\}\rho_{n} \,,
\end{align}
which evaluates to
\begin{align}
	(a^-\{f\}\rho_{n})(x^{(n-1)})
	&= \sum_{j=1}^n
	\sum_{\beta_1\leq \dots \leq \beta_n} \left(\int_\SpaceOfMotion f(y)\, u_{\beta_j}(y) \,dy\right)
	\tilde c_{\beta_1,\dots,\beta_n} \SymmOperator_{n-1} (u_{\setminus \{\beta_j\} })(x^{(n-1)}) \notag \\
	&=\sum_{j=1}^n  \int_\SpaceOfMotion f(y) \rho_n(x_1^{(n-1)},\dots,x_{j-1}^{(n-1)}, y, x_{j+1}^{(n-1)},\dots,x_{n-1}^{(n-1)})dy \notag \\
	&= n\int_\SpaceOfMotion f(y)  \rho_{n}\left(x^{(n-1)}, y\right) \, dy. \label{def:densityaj-2}
\end{align}

\subsection{Conserving single-particle operators}
\label{app:conervsingleoperator}

We prove that the operator $\mathcal{A}=\mathcal{A}^n$ defined in \cref{eq:consvPartMulti} fulfills \cref{eq:singPartExpansion}:
\begin{align}
	\mathcal{A} v &= \sum_{j=1}^n \mathcal{A}_j \SymmOperator_n  \left( v_1\otimes \dots \otimes v_n\right) \notag \\
	& = \SymmOperator_n \sum_{j=1}^n  (A v_j)\otimes v_{\setminus \{j\}} \notag \\
	&\stackrel{\eqref{eq:expansionHelement}}{=}
	\SymmOperator_n \sum_{j=1}^n \left(  \sum_{\alpha} \langle u_\alpha^*, A v_j \rangle u_\alpha\otimes v_{\setminus \{j\}}\right) \notag \\
	& \stackrel{\eqref{eq:resunity}}{=} \SymmOperator_n \sum_{j=1}^n \left(  \sum_{\alpha,\beta} \langle u_\alpha^*, A u_\beta\rangle \langle u_\beta^*, v_j\rangle   u_\alpha\otimes v_{\setminus \{j\}}\right) \notag \\
	&=
	\sum_{\alpha,\beta} \langle u_\alpha^*, A u_\beta\rangle \,\, \SymmOperator_n   \sum_{j=1}^n \langle u_\beta^*, v_j\rangle u_\alpha \otimes v_{\setminus \{j\}} \notag \\
	&\stackrel{\eqref{eq:apam_gen}}{=}
	\sum_{\alpha,\beta} \langle u_\alpha^*, A u_\beta\rangle  \aps{u_\alpha} \ams{u_\beta^*} v.
\end{align}

\subsection{Conserving two-particle operators}
\label{app:conervtwooperator}

Per definition \eqref{def:B_ij} it holds
\begin{equation}
  \mathcal{B} v  = \sum_{1 \leq i < j \leq n} \mathcal{B}_{ij} v
  =\SymmOperator_n\sum_{1 \leq i < j \leq n} B (v_i \otimes v_j) \otimes v_{\setminus \{i,j\}}.
\end{equation}
The representation of $B (v_i \otimes v_j)$ in terms of the symmetrized 2-particle tensor basis reads
[\cref{basis_representation}]:
\begin{equation}
B (v_i \otimes v_j) = \sum_{\alpha \leq \beta} \langle \SymmOperator_{2} (u_\alpha \otimes u_\beta)^*, B(v_i \otimes v_j) \rangle
\, \SymmOperator_{2}(u_\alpha \otimes u_\beta).
\end{equation}
Together this gives
\begin{align}
	\mathcal{B} v & = \SymmOperator_{n} \sum_{1 \leq i < j \leq n}\sum_{\alpha \leq \beta} \langle \SymmOperator_{2} (u_\alpha \otimes u_\beta)^*, B(v_i \otimes v_j) \rangle    \, \SymmOperator_{2}(u_\alpha \otimes u_\beta) \otimes v_{\setminus \{i,j\}}
	\notag \\
	&   = \sum_{1 \leq i < j \leq n}\sum_{\alpha \leq \beta} \langle \SymmOperator_{2} (u_\alpha \otimes u_\beta)^*, B(v_i \otimes v_j) \rangle    \, \SymmOperator_{n} (u_\alpha \otimes u_\beta \otimes v_{\setminus \{i,j\}}) \notag \\
	& \stackrel{\eqref{eq:resunity}}{=}
	\sum_{1 \leq i < j \leq n}\sum_{\substack{\alpha \leq \beta \\ \gamma \leq \delta}} \langle \SymmOperator_{2} (u_\alpha \otimes u_\beta)^*, B \SymmOperator_{2} (u_\gamma \otimes u_\delta)\rangle \langle \SymmOperator_{2} (u_\gamma \otimes u_\delta)^*, \SymmOperator_{2} (v_i \otimes v_j) \rangle \notag \\
	& \hspace{7em} \times \SymmOperator_{n} (u_\alpha \otimes u_\beta \otimes v_{\setminus \{i,j\}}) ,
\end{align}
and after interchanging dualization and symmetrization using \cref{eq:symm-duals}:
\begin{align} \label{Bv}
	\mathcal{B} v
	&= \sum_{1 \leq i < j \leq n}
	\sum_{\substack{\alpha \leq \beta \\ \gamma \leq \delta}}  \frac{1}{s_{\alpha\beta} s_{\gamma\delta}}\langle \SymmOperator_{2} (u^*_\alpha \otimes u^*_\beta), B \SymmOperator_{2} (u_\gamma \otimes u_\delta)\rangle  \langle \SymmOperator_{2} (u^*_\gamma \otimes u^*_\delta), \SymmOperator_{2} (v_i \otimes v_j) \rangle  \nonumber \\
	& \hspace{7em} \times \SymmOperator_{n} (u_\alpha \otimes u_\beta \otimes v_{\setminus \{i,j\}}) \,,
\end{align}
where the symmetrization $\SymmOperator_{2}$ in the argument of the symmetric operator $B$ may be omitted.
Moreover, we note that for $v \in \SymmOperator_n H^{\otimes n}$ it holds
\begin{align}
	\frac{1}{2} \aps{u_\alpha} \aps{u_\beta} & \ams{u_\gamma^*} \ams{u_\delta^*} \, v \nonumber \\
	&\stackrel{\eqref{eq:apapamam_gen}}{=}    \frac{1}{2} \sum_{\substack{i,j=1 \\ j \neq i}}^n \langle u_\delta^*,v_i \rangle   \langle u_\gamma^*,v_j \rangle  \SymmOperator_{n} (u_\alpha \otimes u_\beta \otimes v_{\setminus \{i,j\}}) \nonumber \\
	& =  \frac{1}{2} \sum_{\substack{i,j=1 \\ j \neq i}}^n \langle u_\gamma^*\otimes u_\delta^*, v_i \otimes v_j \rangle
	\SymmOperator_{n} (u_\alpha \otimes u_\beta \otimes v_{\setminus \{i,j\}}) \nonumber \\
	& =  \sum_{1 \leq i < j \leq n} \langle u_\gamma^*\otimes u_\delta^*, \SymmOperator_{2} (v_i \otimes v_j) \rangle  \SymmOperator_{n} (u_\alpha \otimes u_\beta \otimes v_{\setminus \{i,j\}}) \nonumber \\
	& =  \sum_{1 \leq i < j \leq n} \langle \SymmOperator_{2}(u_\gamma^*\otimes u_\delta^*), \SymmOperator_{2}(v_i \otimes v_j)) \rangle  \SymmOperator_{n} (u_\alpha \otimes u_\beta \otimes v_{\setminus \{i,j\}}). 
\end{align}
where \cref{eq:symm-orth} was used in the last line. Inserting this equality into \cref{Bv}, the operator fulfills
\begin{equation}
	\mathcal{B} v =    \frac{1}{2} \sum_{\substack{\alpha \leq \beta \\ \gamma \leq \delta}}  \frac{1}{s_{\alpha\beta} s_{\gamma\delta}} \langle \SymmOperator_{2} (u_\alpha^* \otimes u_\beta^*), B  (u_\gamma \otimes u_\delta)\rangle \,\, \aps{u_\alpha} \aps{u_\beta} \ams{u_\gamma^*} \ams{u_\delta^*} \, v.
\end{equation}
By means of \cref{s_alphabeta}, and using the symmetry of the summands, we finally obtain
\begin{equation}
	\mathcal{B} v =    \frac{1}{2} \sum_{\alpha,\beta,\gamma,\delta}   \langle u_\alpha^* \otimes u_\beta^*, B  (u_\gamma \otimes u_\delta)\rangle \,\, \aps{u_\alpha} \aps{u_\beta} \ams{u_\gamma^*} \ams{u_\delta^*} \, v.
\end{equation}

Note that the need of the factor $1/2$ becomes clear as a consequence of applying the annihilation operator twice; removing first particle $i$ and then $j$ is the same as removing particle $j$ and then $i$. If an operator expansion consists of $n$ annihilation operators, it will require a factor $1/n!$ to account for all the equivalent orderings.

One can verify the equation above directly by expanding the creation and annihilation operators and by contracting the expansions in the non-symmetrized basis:
\begin{align}
		\frac{1}{2} & \sum_{\alpha ,\beta ,\gamma ,\delta} \langle u_\alpha^* \otimes u_\beta^*, B (u_\gamma \otimes u_\delta)\rangle \,\, \aps{u_\alpha} \aps{u_\beta} \ams{u_\gamma^*} \ams{u_\delta^*} v \notag \\
		& \stackrel{\eqref{eq:apapamam_gen}}{=}    \frac{1}{2}\sum_{\alpha ,\beta ,\gamma ,\delta} \langle  u_\alpha^* \otimes u_\beta^*, B (u_\gamma \otimes u_\delta)\rangle \,\,  \SymmOperator_{n}\left(\sum_{\substack{i,j=1 \\ j \neq i}}^n \langle u_\delta^*,v_i \rangle   \langle u_\gamma^*,v_j \rangle (u_\alpha \otimes u_\beta \otimes v_{\setminus \{i,j\}})  \right) \notag \\
		& = \frac{1}{2}
		\SymmOperator_{n} \sum_{\substack{i,j=1 \\ j \neq i}}^n\sum_{\alpha ,\beta ,\gamma ,\delta} \langle u_\alpha^* \otimes u_\beta^*, B(u_\gamma \otimes u_\delta)\rangle \,\,  \langle u_\gamma^* \otimes u_\delta^*, v_i \otimes v_j \rangle    (u_\alpha \otimes u_\beta \otimes v_{\setminus \{i,j\}}) \notag \\
		& \stackrel{\eqref{eq:resunity}}{=} \frac{1}{2} \SymmOperator_{n} \sum_{\substack{i,j=1 \\ j \neq i}}^n B (v_i \otimes v_j) \otimes v_{\setminus \{i,j\}} \notag \\
		& \stackrel{\eqref{def:B_ij}}{=} \sum_{1 \leq i < j \leq n} \mathcal{B}_{ij} v \notag \\
		& =  \mathcal{B} v.
\end{align}

\subsection{Reaction operators}
\label{app:reactionoperators}
The reaction operators
\begin{subequations}
\begin{align}
	\ReactionOperator^{(k)}&= \frac{1}{k!}\sum_{\substack{\alpha_1, \dots , \alpha_k \\ \beta_1, \dots , \beta_k}} \left\langle \bigotimes_{i=1}^k u_{\alpha_i}^* , \Lambda^{(k)} \bigotimes_{j=1}^k u_{\beta_j} \right\rangle \,\,  \prod_{i=1}^{k} \aps{u_{\alpha_i}}\prod_{j=1}^{k} \ams{u_{\beta_j}^*},\\
	\ReactionOperator^{(k,l)} &= \frac{1}{k!}\sum_{\substack{ \alpha_1 ,\dots ,\alpha_l \\ \beta_1 , \dots , \beta_k}} \left\langle   \bigotimes_{i=1}^l u_{\alpha_i}^* , \Lambda^{(k,l)} \bigotimes_{j=1}^k u_{\beta_j} \right\rangle \prod_{i=1}^{l} \aps{u_{\alpha_i}}\prod_{j=1}^{k} \ams{u_{\beta_j}^*}
\end{align}
\end{subequations}
can also be written in terms of their symmetric basis:
\begin{subequations}
\begin{align}
	\ReactionOperator^{(k)}&= \frac{1}{k!}\sum_{\substack{\alpha_1 \leq \dots \leq \alpha_k \\ \beta_1 \leq \dots \leq \beta_k}} \frac{1}{s_{\beta_1,\dots,\beta_k}}\left\langle \SymmOperator_k\left(\bigotimes_{i=1}^k u_{\alpha_i}\right)^* , \Lambda^{(k)} \bigotimes_{j=1}^k u_{\beta_j} \right\rangle \,\,  \prod_{i=1}^{k} \aps{u_{\alpha_i}}\prod_{j=1}^{k} \ams{u_{\beta_j}^*},\\
	\ReactionOperator^{(k,l)} &= \frac{1}{k!}\sum_{\substack{\alpha_1 \leq \dots \leq \alpha_l \\ \beta_1 \leq \dots \leq \beta_k}} \frac{1}{s_{\beta_1,\dots,\beta_k}}  \left\langle \SymmOperator_k\left( \bigotimes_{i=1}^l u_{\alpha_i}\right)^* , \Lambda^{(k,l)} \bigotimes_{j=1}^k u_{\beta_j} \right\rangle \prod_{i=1}^{l} \aps{u_{\alpha_i}}\prod_{j=1}^{k} \ams{u_{\beta_j}^*},
\end{align}
\end{subequations}
where $s_{\beta_1,\dots,\beta_n}$ is defined on \cref{eq:symm-coeff}. The proofs of these expansions work analogously to the proof in \cref{app:conervtwooperator}. The coefficients $s_{\beta_1,\dots,\beta_k}$ and the coefficients $s_{\alpha_1,\dots,\alpha_k}$ (or $s_{\alpha_1,\dots,\alpha_l}$) that appear when commuting symmetrization with duality compensate for the duplicate terms in the non-symmetrized expansion.

\section{Local conservation of probabilities and dissipativity of the CDME operator} \label{sec:dissipative}

\rrev{%
The dissipativity of the reaction operator is closely related to the local conservation of probability, expressed by \cref{eq:conservation}, which we will prove first. 
This can be done for each reaction separately, such that we can restrict to the case of a single reaction with fixed stoichiometric coefficients $k$ and $l$. 
Then each ``column'' $m$ of the reaction operator $\ReactionOperator = (\mathcal{Q}_{nm})$ has at most two non-zero entries (see \cref{eq:Qmatrix}) and the column sum reduces to
\begin{align}\label{C1}
 \sum_{n=0}^\infty \mathcal{J}_{nm}(\rho_m) &= \mathcal{J}_{m-k+l,m}(\rho_m) + \mathcal{J}_{mm}(\rho_m) \notag \\
  &= \int_{\SpaceOfMotion^{m-k+l}} \, \ReactionOperator^{(k,l)}\rho_m  dx^{(m-k+l)}
   - \int_{\SpaceOfMotion^m} \, \ReactionOperator^{(k)}\rho_m  dx^{(m)}
\end{align}
for $m \geq k$ and dropping the subscript $r$ on the reaction operators.
(For $m < k$, the reaction does not occur and $\mathcal{J}_{nm}(\rho_m) = 0$, irrespective of the value of $n$.)
 For the sum in \eqref{C1} to be zero it has to hold 
\begin{equation}\label{inflow_outflow}
	\int_{\SpaceOfMotion^m}(\ReactionOperator^{(k)}\rho_m)(x^{(m)}) dx^{(m)} = \int_{\SpaceOfMotion^{m-k+l}}(\ReactionOperator^{(k,l)}\rho_m)(x^{(m-k+l)}) dx^{(m-k+l)} \,,
\end{equation}
for $\rho_m\in  \SymmOperator_m \FunctionSpace^{\otimes m}$,} which means that the considered reaction induces the probability outflow from the $m$-particle space given on the l.h.s., which is balanced by the inflow into the space with $m-k+l$ particles on the r.h.s.

We here give the proof of \cref{inflow_outflow} for the case $k=2$ and $l=1$, i.e., the reaction $\Species + \Species  \to \Species$; the general case for arbitrary $k,l$ follows analogously.
The reaction operators read [\cref{Rk,R10}]:
\begin{align}
	\ReactionOperator^{(2)}&= \frac{1}{2}\sum_{\substack{\alpha_1, \alpha_2\\ \beta_1,\beta_2}} \langle u_{\alpha_1}^* \otimes u_{\alpha_2}^*, \Lambda^{(2)}( u_{\beta_1} \otimes u_{\beta_2} )\rangle \,\, \aps{u_{\alpha_1}} \aps{u_{\alpha_2}} \ams{u_{\beta_1}^*} \ams{u_{\beta_2}^*},  \label{eq:appC_R2} \\
	\ReactionOperator^{(2,1)} &= \frac{1}{2}\sum_{\alpha} \sum_{\beta_1 ,\beta_2} \langle u_\alpha^*, \Lambda^{(2, 1)} (u_{\beta_1} \otimes u_{\beta_2})  \rangle \aps{u_\alpha} \ams{u_{\beta_1}^*} \ams{u_{\beta_2}^*}, \label{eq:appC_R21}
\end{align}
with the coefficients given in terms of the same reaction rate function $\lambda(x^{(1)},x^{(2)})$ as [\cref{eq:propOperators_k,eq:propOperators_kl}]:
\begin{align}
\left(\Lambda^{(2)}( u_{\beta_1} \otimes u_{\beta_2})\right)(x^{(2)}) &=
 (u_{\beta_1} \otimes u_{\beta_2})(x^{(2)}) \int_{\SpaceOfMotion^1} \lambda(x^{(1)},x^{(2)}) dx^{(1)},
\\
\left(\Lambda^{(2, 1)} (u_{\beta_1} \otimes u_{\beta_2})\right) (x^{(1)}) &=
\int_{\SpaceOfMotion^2}( u_{\beta_1} \otimes u_{\beta_2})(x^{(2)}) \lambda(x^{(1)},x^{(2)}) dx^{(2)}.
\end{align}
Integrating the remaining degrees of freedom of each equation and using Fubini's theorem, we obtain the relation
\begin{align}
	\int_{\SpaceOfMotion^2} \left(\Lambda^{(2)}( u_{\beta_1} \otimes u_{\beta_2})\right)(x^{(2)}) dx^{(2)} = \int_{\SpaceOfMotion^1} \left(\Lambda^{(2, 1)} (u_{\beta_1} \otimes u_{\beta_2})\right) (x^{(1)}) dx^{(1)}. \label{eq:appC_equalInts}
\end{align}
With this, we conclude \cref{inflow_outflow} as follows:
\begingroup
\allowdisplaybreaks
\begin{align}
\mathrlap{\int_{\SpaceOfMotion^m} \ReactionOperator^{(2)} \rho_m \, dx^{(m)}} \hspace{.5em} \notag \\
	&\stackrel{\mathclap{\eqref{eq:appC_R2}}}{=} \int_{\SpaceOfMotion^m} \frac{1}{2}\sum_{\substack{\alpha_1, \alpha_2\\ \beta_1,\beta_2}} \langle u_{\alpha_1}^* \otimes u_{\alpha_2}^*, \Lambda^{(2)}( u_{\beta_1} \otimes u_{\beta_2} )\rangle \,\, \aps{u_{\alpha_1}} \aps{u_{\alpha_2}} \ams{u_{\beta_1}^*} \ams{u_{\beta_2}^*} \rho_m dx^{(m)} \,, \notag  \\
	\intertext{inserting the definition of the creation operator [\eqref{def:longsymaj+}],}
	&=
	\int_{\SpaceOfMotion^m} \frac{1}{2}\sum_{\substack{\alpha_1, \alpha_2\\ \beta_1,\beta_2}} \langle u_{\alpha_1}^* \otimes u_{\alpha_2}^*, \Lambda^{(2)}( u_{\beta_1} \otimes u_{\beta_2} )\rangle \,\, \SymmOperator_{n} (u_{\alpha_1} \otimes u_{\alpha_2} \otimes \ams{u_{\beta_1}^*} \ams{u_{\beta_2}^*} \rho_m)dx^{(m)} \,, \notag  \\
\intertext{omitting the symmetrization under the integral,}
	&=
	\int_{\SpaceOfMotion^m} \frac{1}{2}\sum_{\substack{\alpha_1, \alpha_2\\ \beta_1,\beta_2}} \langle u_{\alpha_1}^* \otimes u_{\alpha_2}^*, \Lambda^{(2)}( u_{\beta_1} \otimes u_{\beta_2} )\rangle \,\,  u_{\alpha_1} \otimes u_{\alpha_2} \otimes \ams{u_{\beta_1}^*} \ams{u_{\beta_2}^*} \rho_mdx^{(m)}\,, \notag  \\
	\intertext{carrying out the sums over $\alpha_1, \alpha_2$ upon recognizing the expansion of $\Lambda^{(2)}( u_{\beta_1} \otimes u_{\beta_2})$ in the 2-particle basis [\cref{eq:v_basis2}],}
	&= \int_{\SpaceOfMotion^m} \frac{1}{2}\sum_{\beta_1,\beta_2} \Lambda^{(2)}( u_{\beta_1} \otimes u_{\beta_2} ) \otimes \ams{u_{\beta_1}^*} \ams{u_{\beta_2}^*} \rho_m dx^{(m)} \,, \notag  \\
	\intertext{interchanging the sums and integrals,}
	&= \frac{1}{2}\sum_{\beta_1,\beta_2} \int_{\SpaceOfMotion^{2}} \Lambda^{(2)}( u_{\beta_1} \otimes u_{\beta_2} ) dx^{(2)} \int_{\SpaceOfMotion^{n-2}}  \ams{u_{\beta_1}^*} \ams{u_{\beta_2}^*} \rho_m dx^{(m-2)} \,, \notag \\
	\intertext{making use of the identity \eqref{eq:appC_equalInts},}
	&= \frac{1}{2}\sum_{\beta_1,\beta_2} \int_{\SpaceOfMotion} \Lambda^{(2,1)}( u_{\beta_1} \otimes u_{\beta_2} ) dx^{(1)} \int_{\SpaceOfMotion^m}  \ams{u_{\beta_1}^*} \ams{u_{\beta_2}^*} \rho_m dx^{(m-2)} \,, \notag \\
	\intertext{and finally expanding $\Lambda^{(2,1)}( u_{\beta_1} \otimes u_{\beta_2} )$ in the 1-particle basis [\cref{eq:v_basis2}],}
	&= \frac{1}{2}\sum_{ \beta_1,\beta_2} \int_{\SpaceOfMotion} \sum_\alpha \langle u_\alpha^*,\Lambda^{(2, 1)} (u_{\beta_1} \otimes u_{\beta_2}) \rangle \aps{u_{\alpha}} dx^{(1)}\int_{\SpaceOfMotion^{n-2}}  \ams{u_{\beta_1}^*} \ams{u_{\beta_2}^*} \rho_m dx^{(m-2)} \notag \\
	&= \int_{\SpaceOfMotion^{n-1}}  \frac{1}{2}\sum_{\substack{\alpha\\ \beta_1,\beta_2}}  \langle u_\alpha^*,\Lambda^{(2, 1)} (u_{\beta_1} \otimes u_{\beta_2}) \rangle \aps{u_{\alpha}} \ams{u_{\beta_1}^*} \ams{u_{\beta_2}^*} \rho_m  dx^{(n-1)} \notag \\
	&\stackrel{\mathclap{\eqref{eq:appC_R21}}}{=} \int_{\SpaceOfMotion^{n-1}}\ReactionOperator^{(2,1)}\rho_m dx^{(n-1)}.
\end{align}
\endgroup
\rrev{This implies \eqref{inflow_outflow} for the considered reaction. Using equivalent argumentation for other reactions and combining the results delivers \cref{eq:conservation}.}

\rrev{
For the dissipativity of the reaction operator $\ReactionOperator$ one has to show that
\begin{equation}
  \|\mu \rho - \ReactionOperator \rho\| \geq \mu \|\rho\| \quad \text{for all $\mu >0$ and $\rho \in \dom(\ReactionOperator)$}.
\end{equation} 
Here, $\dom(\ReactionOperator) \subset \Fockspace$ denotes the domain of $\ReactionOperator$,
which is taken such that $\rho \in \dom(\ReactionOperator)$ satisfies $\| \ReactionOperator \rho \| < \infty$ and
$\sum_{n=0}^\infty |\mathcal{J}_{nm}(\rho_m)| < \infty$ for each $m \in \mathbb{N}_0$.
The last condition together with \cref{eq:conservation} implies the relation
\begin{multline}
  \label{eq:Rrho-zero}
 \sum_{n=0}^\infty \int_{\SpaceOfMotion^n} (\ReactionOperator \rho)_n dx^{(n)}
 = \sum_{n=0}^\infty \int_{\SpaceOfMotion^n} \sum_{m=0}^\infty (\AltReactionOperator_{nm} \rho_m) dx^{(n)}
 = \sum_{m=0}^\infty \left(\sum_{n=0}^\infty \mathcal{J}_{nm}(\rho_m) \right) \stackrel{\eqref{eq:conservation}}{=} 0,
\end{multline}
noting that only finitely many of the $\mathcal{Q}_{nm}$ are non-zero for fixed $n$ (see \cref{eq:Qmatrix}), which allows us to interchange the $m$-sum and the integral over $\SpaceOfMotion^n$ and finally the order of summation.
With this relation, it is straightforward to infer the dissipativity of $\ReactionOperator$:
\begin{align}
  \|\mu \rho - \ReactionOperator \rho\| &= \sum_{n=0}^\infty \|\mu \rho_n - (\ReactionOperator \rho)_n \|_n \notag \\
  &\geq \Big|\sum_{n=0}^\infty \int_{\SpaceOfMotion^n}[\mu \rho_n -(\ReactionOperator \rho)_n] \, dx^{(n)} \Big| \notag \\
  &= \Big|\sum_{n=0}^\infty \int_{\SpaceOfMotion^n}\mu \rho_n dx^{(n)} - \sum_{n=0}^\infty \int_{\SpaceOfMotion^n} (\ReactionOperator \rho)_n \, dx^{(n)} \Big| \notag \\
  &\hspace{-1ex}\stackrel{\eqref{eq:Rrho-zero}}{=} \mu \|\rho\|, \label{dissipativity}
\end{align}
making use of the triangle inequality in the second line and that both sums are absolutely convergent in the last-but-one step.
}

\rrev{Finally, the calculation in \eqref{dissipativity} can directly be transferred to the reaction--diffusion operator $\mathcal{D}+\ReactionOperator$, noticing that due to the reflecting boundary conditions at the boundary of $\mathbb{X}$, it holds
	\begin{equation}\label{eq:D=0}
	\int_{\SpaceOfMotion^n} (\mathcal{D}\rho_n)(x^{(n)}) dx^{(n)} =0
	\end{equation}
	for any $n\in \mathbb{N}$. In total, we obtain the dissipativity of the CDME operator $\mathcal{A}=\ReactionOperator + \mathcal{D}$. }

\section{Spatial discretization}
\label{app:spatialdiscretization}

The crucial step in performing the spatial discretization is to consistently define the respective diffusion and reaction operators, which are expressed in terms of creation and annihilation operators, acting on \rev{the basis elements of the copy number representation}. 
We lay a basis for doing so in the following.

\subsection{Creation and annihilation operators acting \rev{on a discrete space}}\label{app:ann-ops}

The action of the creation and annihilation operators on the Galerkin-projected Fock space $\hat F$ is particularly transparent for basis elements of the copy number representation. Using the short-hand notation \eqref{copynumber}, it holds
\begin{align}
	a_k^+ \ket{N_1,\dots,N_M} &= \hphantom{N_k} \ket{N_1,\dots,N_k + 1, \dots, N_M} ,\label{eq:cre-ops} \\
	a_k^- \ket{N_1,\dots,N_M} &= N_k \ket{N_1,\dots,N_k - 1, \dots, N_M}. \label{eq:ann-ops1}
\end{align}
Relation \eqref{eq:cre-ops} follows directly from the definition of the operator $a_k^+$.
In order to prove the second relation, \cref{eq:ann-ops1}, we make the following observation.
From the commutator relation $\left[a^- \{f\}, a^+ \{w\}\right] = \left\langle f,w \right\rangle  \Id$, see \cref{eq:a+a-CommutRels}, we get
\begin{equation}
	a^-_i a^+_j = a^+_j a^-_i + \left\langle \xi_i^*,\xi_j \right\rangle \Id = a^+_j a^-_i + \delta_{ij} \Id,
\end{equation}
where $\delta_{ij}$  is the Kronecker delta. More generally, for $n\geq 1$, we have
\begin{align}
	a_j^- (a_i^+)^n &= a_i^+a_j^-(a_i^+)^{n-1} + \delta_{ij}(a_i^+)^{n-1}\nonumber\\
	&=  (a_i^+)^2a_j^-(a_i^+)^{n-2} + 2\delta_{ij}(a_i^+)^{n-1}\nonumber\\
	&=  \ldots \nonumber\\
	&=  (a_i^+)^na_j^- + n\delta_{ij}(a_i^+)^{n-1},\nonumber
\end{align}
and thus
\begin{equation}
	\left[ a_j^-,(a_i^+)^n \right] = n \delta_{ij}(a_i^+)^{(n-1)}.
\end{equation}
We can use this formula to calculate the following identities:
\begin{align}
	\lefteqn{a_k^- (a_1^+)^{N_1}\ldots(a_M^+)^{N_M} } \qquad \nonumber \\
	& = (a_1^+)^{N_1}\ldots(a_M^+)^{N_M} a_k^- + \sum_{i=1}^M N_i \delta_{ik}(a_1^+)^{N_1}\ldots(a_i^+)^{N_i -1}\ldots(a_M^+)^{N_M} \nonumber \\
	&=	(a_1^+)^{N_1}\ldots(a_M^+)^{N_M} a_k^- + N_k (a_1^+)^{N_1}\ldots(a_k^+)^{N_k -1}\ldots(a_M^+)^{N_M},
\end{align}
which, applied to $\rho_{vac}$, yields \cref{eq:ann-ops1}.
By repeated application of \cref{eq:ann-ops1} if follows directly that
\begin{equation}
	a_k^- a_l^- \ket{N_1,\dots,N_M} = N_k N_l \ket{N_1, \ldots, N_k -1, \ldots, N_l -1, \ldots, N_M}
	\label{eq:ann-ops}
\end{equation}
for $k\neq l$, and
\begin{equation}
	a_k^- a_k^- \ket{N_1,\dots,N_M} = N_k (N_k-1) \ket{N_1, \ldots, N_k -2,  \ldots, N_M},
\end{equation}
which will be used in the next section for deriving the projected reaction operators of mutual annihilation.

\subsection{Diffusion and reaction operators}\label{app:diff-reac-ops}

For the calculation of the diffusion and reaction operators we need to apply the properties of the creation and annihilation operators given in  \cref{app:ann-ops} and subsequently shift indices.
In order to simplify notation in this index shift, we define $p_{N_1,\ldots, N_M}:=0$ for all $(N_1,\dots,N_M) \notin  \mathbb{N}_0^M$ (especially for negative indices).

\paragraph*{Diffusion.}

For the projected diffusion operator $\hat{\mathcal{D}}$ given in \cref{D_projected} and specified in \cref{D_projected2} we find, using the copy number representation  \eqref{Eq:Galerkin-B} as well as the properties \eqref{eq:cre-ops} and \eqref{eq:ann-ops1},
\begin{align} \label{eq:Dhat}
	\hat{\mathcal{D}} \hat{\rho} =&   	\sum_{N_1,\dots,N_M}
	p_{N_1,\ldots, N_M} \sum_{i,j}d_{ij} a_i^+ a_j^-
	\ket{N_1,\dots,N_M} \nonumber \\
	= &	\sum_{N_1,\dots,N_M}
	p_{N_1,\ldots, N_M} \sum_{\substack{i,j \\ i\neq j}}d_{ij}
	N_j\cdot
	\ket{N_1,\dots,N_i+1,\ldots,N_j-1,\ldots,N_M} \nonumber \\
	& \quad + \sum_{N_1,\dots,N_M}
	p_{N_1,\ldots, N_M} \sum_{i}d_{ii}  N_i\cdot
	\ket{N_1,\ldots,N_M} \nonumber \\
	=&   
	\sum_{N_1,\dots,N_M} \sum_{\substack{i,j \\ i\neq j}}
	p_{N_1,\ldots, N_i-1, \ldots, N_j+1, \ldots,  N_M} d_{ij}
	(N_j+1)\cdot
	\ket{N_1,\ldots,N_M} \nonumber \\
	& \quad + \sum_{N_1,\dots,N_M} \sum_{i}
	p_{N_1,\ldots, N_M}d_{ii}  N_i\cdot \ket{N_1,\ldots,N_M}.
\end{align}

\paragraph*{Degradation.}

Analogously, we obtain for the conserving one-particle reaction operator $\hat{\ReactionOperator}_d^{(1)}=\sum_{i,j} \langle \xi_i^*, \Lambda^{(1)}_d \xi_j\rangle a_i^+ a_j^-$ of degradation $\Species\to \emptyset$ by using definition \eqref{def:lambda}:
\begin{align} \label{eq:Rd1_hat}
	\hat{\ReactionOperator}_d^{(1)} \hat\rho &=
	\sum_{N_1,\dots,N_M}
	p_{N_1,\ldots, N_M} \sum_{i,j}\Propensity_d^{ij}a_i^+ a_j^-
	\ket{N_1,\ldots,N_M} \nonumber \\
	&= \sum_{N_1,\dots,N_M}
	p_{N_1,\ldots, N_M} \sum_{\substack{i,j\\ i\neq j}}\Propensity_d^{ij}
	N_j\cdot
	\ket{N_1,\ldots,N_i+1,\ldots,N_j-1,\ldots,N_M} \nonumber \\
	& \qquad +\sum_{N_1,\dots,N_M}
	p_{N_1,\ldots, N_M} \sum_{i}\Propensity_d^{ii}  N_i\cdot
	\ket{N_1\ldots,N_M} \nonumber \\
	&=
	\sum_{N_1,\dots,N_M} \sum_{\substack{i,j \\i\neq j}}
	p_{N_1,\ldots, N_i-1, \ldots, N_j+1, \ldots,  N_M} \Propensity_d^{ij}
	(N_j+1)\cdot
	\ket{N_1,\ldots,N_M} \nonumber \\
	& \qquad +\sum_{N_1,\dots,N_M} \sum_{i}
	p_{N_1,\ldots, N_M}\Propensity_d^{ii}  N_i\cdot
	\ket{N_1,\ldots,N_M},
\end{align}
and for the non-conserving one-particle reaction operator $\hat{\ReactionOperator}_d^{(1,0)}=\sum_i \langle 1, \Lambda^{(1,0)}_d \xi_i\rangle a_i^-$ of degradation, by means of definition \eqref{def:lambda}:
\begin{align} \label{eq:Rd10_hat}
	\hat{\ReactionOperator}_d^{(1,0)} \hat{\rho}
	&= \sum_{N_1,\dots,N_M}  p_{N_1,\ldots, N_M} \sum_i\Propensity_d^i  \, a^-_i \ket{N_1,\ldots,N_M} \nonumber \\
	&= \sum_{N_1,\dots,N_M} p_{N_1,\ldots, N_M} \, \sum_i\Propensity_d^i  N_i\cdot \ket{N_1,\ldots,N_i-1,\ldots,N_M} \nonumber \\
	&= \sum_{N_1,\dots,N_M} \sum_i p_{N_1,\ldots, N_i+1, \ldots, N_M} \,\Propensity_d^i  (N_i+1)\cdot \ket{N_1,\ldots,N_M}.
\end{align}

\paragraph*{Creation.}
For the creation reaction $\emptyset \to I $ the conserving part is given by
\begin{equation} \label{eq:Rc0_hat}
	\hat{\ReactionOperator}_c^{(0)} \hat\rho  = \sum_{N_1,\dots,N_M} p_{N_1,\ldots, N_M} \, \Propensity_c \ket{N_1,\ldots,N_M},
\end{equation}
and the nonconserving one-particle reaction operator acts as:
\begin{align} \label{eq:Rc01_hat}
	\hat{\ReactionOperator}_c^{(0,1)} \hat\rho &=
	\sum_{N_1,\dots,N_M}
	p_{N_1,\ldots, N_M} \, \sum_i\Propensity_c^i  a^+_i
	\ket{N_1,\ldots,N_M}\nonumber\\
	&=\sum_{N_1,\dots,N_M}
	p_{N_1,\ldots, N_M} \,\sum_i \Propensity_c^i
	\ket{N_1,\ldots,N_i +1,\ldots,N_M}\nonumber\\
	&= \sum_{N_1,\dots,N_M} \sum_i
	p_{N_1,\ldots, N_i -1,\ldots,  N_M}\Propensity_c^i   \,
	\ket{N_1,\ldots,N_M},
\end{align}
where we applied definition \eqref{def:lambda_01}.

\paragraph*{Mutual annihilation.}

For the reaction $\Species + \Species  \to \emptyset$ of mutual annihilation, the projected conserving  operator  $\hat{\ReactionOperator}^{(2)}$ is given by
\begin{equation}	\hat{\ReactionOperator}^{(2)} = \frac{1}{2}\sum_{\substack{i, j \\ k,l}} \langle \xi^*_i \otimes \xi^*_j, \Lambda^{(2)} (\xi_k \otimes \xi_l) \rangle  a_i^+ a_j^+ a_k^- a_l^-
\end{equation}
see \cref{R2}. For the rescaled indicator functions given in \cref{rescaled_indicator}, the factor
\begin{equation}\label{factor}
	\langle \xi^*_i \otimes \xi^*_j, \Lambda^{(2)} (\xi_k \otimes \xi_l) \rangle = \int_{\SpaceOfMotion^2}  (\xi^*_i \otimes \xi^*_j)(x^{(2)})  \Propensity(x^{(2)}) (\xi_k \otimes \xi_l)(x^{(2)}) dx^{(2)}
\end{equation}
is for $k=i$ and $l=j$ equal to $\lambda^{ij}$ defined in \cref{lambda_mutual},  and zero otherwise.

We thus obtain
\begin{align} \label{eq:R2_hat}
	\hat{\ReactionOperator}^{(2)} \hat\rho &=
	\frac{1}{2} \sum_{N_1,\dots,N_M}
	\sum_{i, j}
	p_{N_1,\ldots, N_M} \Propensity^{ij}  \,
	a^+_i  a^+_j a^-_i a^-_j
	\ket{N_1,\ldots,N_M}
	\nonumber\\
	&=\frac{1}{2}
	\sum_{N_1,\dots,N_M}
	\Bigg[
	\sum_{i}
	p_{N_1,\ldots, N_M} \Propensity^{ii} N_i(N_i-1)\,
	\ket{N_1,\ldots,N_M} \nonumber \\
	& \hspace{7em} + \sum_{i\neq j}
	p_{N_1,\ldots, N_M} \Propensity^{ij} N_i N_j \,
	\ket{N_1,\ldots,N_M}
	\Bigg] \nonumber\\
	&=
	\sum_{N_1,\dots,N_M}
	\Bigg[
	\sum_{i}
	p_{N_1,\ldots, N_M} \frac{1}{2}\Propensity^{ii} N_i(N_i-1)\,
	\ket{N_1,\ldots,N_M} \nonumber \\
	& \hspace{7em} + \sum_{i< j}
	p_{N_1,\ldots, N_M} \Propensity^{ij} N_i N_j \,
	\ket{N_1,\ldots,N_M}
	\Bigg],
\end{align}
where the last equality follows from the fact that the reaction rate function $\lambda(x^{(2)})$ is symmetric, which implies $\lambda^{ij}=\lambda^{ji}$.

For the nonconserving part it holds
\begin{equation}	\hat{\ReactionOperator}^{(2,0)} = \frac{1}{2}\sum_{i, j } \langle 1, \Lambda^{(2,0)} (\xi_i \otimes \xi_j) \rangle  a_i^- a_j^- ,
\end{equation}
where
\begin{align}
	\langle 1, \Lambda^{(2,0)} (\xi_i \otimes \xi_j) \rangle &= \Lambda^{(2,0)} (\xi_i \otimes \xi_j)  \nonumber \\
	&= \int_{\SpaceOfMotion^2} \Propensity(x^{(2)}) (\xi_i \otimes \xi_j)(x^{(2)}) dx^{(2)} \nonumber \\
	&= \Propensity^{ij}
\end{align}
again for  $\Propensity^{ij}$ defined in \cref{lambda_mutual}, such that
\begin{align} \label{eq:R20_hat}
	\hat{\ReactionOperator}^{(2,0)} \hat\rho
	&= \frac{1}{2}
	\sum_{N_1,\dots,N_M}
	\sum_{i,j}
	p_{N_1,\ldots, N_M} \Propensity^{ij}\, a^-_i a^-_j
	\ket{N_1,\ldots,N_M}
	\nonumber\\
	&= \frac{1}{2}
	\sum_{N_1,\dots,N_M} \Bigg[
	\sum_{i}
	p_{N_1,\ldots, N_M} \Propensity^{ii} N_i(N_i-1) \,
	\ket{N_1,\ldots,N_i-2,\ldots,N_M}\nonumber\\
	& \hspace{5em} +\sum_{i\neq j}
	p_{N_1,\ldots, N_M} \Propensity^{ij} N_i N_j \,
	\ket{N_1,\ldots,N_i -1,\ldots,N_j -1,\ldots,N_M}\Bigg]
	\nonumber\\
	&=
	\sum_{N_1,\dots,N_M} \Bigg[
	\sum_{i}
	p_{N_1,\ldots, N_i+2,\ldots, N_M} \frac{1}{2} \Propensity^{ii} (N_i+2)(N_i+1) \,
	\ket{N_1,\ldots,N_M}\nonumber\\
	& \hspace{5em} +\sum_{i<j}
	p_{N_1,\ldots, N_i+1,\ldots, N_j+1, \ldots, N_M} \Propensity^{ij} (N_i+1) (N_j+1) \, 	\ket{N_1,\ldots,N_M}\Bigg].
\end{align}

\bibliography{references}

\end{document}